\documentclass[journal,10pt]{IEEEtran}

\usepackage{}
\usepackage{amsmath}
\usepackage{amsfonts}
\usepackage{mathrsfs}
\usepackage{amssymb}
\usepackage{bm}
\usepackage{cases}
\usepackage{stmaryrd}
\usepackage{mathrsfs}
\usepackage{bbm}
\usepackage[dvips]{graphicx}
\usepackage{CJK}
\usepackage{amsmath}
\usepackage{flushend}
\usepackage{algorithm}
\usepackage{algorithmic}
\usepackage{multirow}
\usepackage{array}
\usepackage{graphicx}
\usepackage{setspace}
\usepackage{subfigure}
\usepackage{epstopdf}
\usepackage{float}
\usepackage{cite}
\usepackage{enumerate}
\usepackage{xcolor}
\usepackage{stfloats}

\newtheorem{remark}{\textbf{Remark}}

\begin{document}

\title{UAV Secure Downlink NOMA Transmissions: A Secure Users Oriented Perspective}
\author{
	Hui-Ming~Wang,~\IEEEmembership{Senior Member,~IEEE}~and
	Xu~Zhang
}

\maketitle

\begin{abstract}
This paper proposes a secure downlink multi-user transmission scheme enabled by a flexible unmanned aerial vehicle base station (UAV-BS) and non-orthogonal multiple access (NOMA). According to their heterogeneous service requirements, multiple legitimate users are categorized as security-required users (SUs) and quality of service (QoS)-required users (QUs), while these QUs can potentially act as internal eavesdroppers which are curious about the secrecy transmissions of SUs. In such a context, our goal is to maximize the achievable minimum secrecy rate among SUs through the joint optimization of user scheduling, power allocation, and trajectory design, subject to the QoS requirements of QUs and the mobility constraint of UAV-BS. Due to the non-convexity of the problem, an efficient iterative algorithm is firstly proposed, based on the alternative optimization (AO) and successive convex approximation (SCA) methods and along with a penalty-based algorithm to deal with the introduced binary integer variables, to obtain a sub-optimal solution. Then, we propose an SUs-oriented low-complexity algorithm by taking advantage of the inherent characteristics of the optimization problem, which can efficiently reduce the computational complexity and can act as a reasonable initial solution for the previous iterative algorithm to achieve better performance. Finally, the superiority of our proposed scheme compared with the conventional orthogonal multiple access (OMA) one is validated by numerical simulation results.
\end{abstract}


\section{Introduction}
Due to their characteristics of flexible deployment and controllable mobility, low acquisition and maintenance costs, high maneuverability, and hovering ability, small-scale unmanned aerial vehicles (UAVs) are promising to act as aerial terminals to support a wide range of civil applications, such as aerial photography, emergency search and rescue, resource exploration and cargo transport. Recently, UAVs have also been considered as aerial platforms in future communication systems \cite{R1,R2,R17}. For instance, UAVs can provide temporary connectivity services for emergency situations without traditional cellular infrastructure coverage due to natural disasters or data traffic offloading in a hotspot area with densely distributed users. Especially under rural and suburban scenarios. it is noted that aerial-to-ground (A2G) line-of-sight (LoS) channels are likely to provide channel superiority compared with terrestrial channels \cite{R18}, which are significantly affected by severe fading and shadowing effects. Moreover, new degrees of freedom (DoFs) introduced by the trajectory design of UAVs can be exploited to facilitate more efficient and reliable transmissions.


Due to the scarcity of public spectrum resources and the explosive growth in data traffic demand of users, one of the significant goals of UAV-involved wireless communications is to improve the spectral efficiency performance. To that end, non-orthogonal multiple access (NOMA) technology \cite{R20} in cellular networks are also promising in UAV communication networks, where the targeted information-bearing signals of multiple users are superimposed for transmission and the efficient multiuser detection technique is exploited at the receiver via successive interference cancellation (SIC). It is noted that based on the positions of ground users, the mobility of UAVs can be exploited to generate channel gain differences among the targeted users in an opportunistic manner, which are useful for NOMA transmissions. As a result, the performance of power-domain NOMA can be effectively improved and there is no need for UAV to be as close to marginal users as possible for service. There have been several works combined with power-domain NOMA and UAV communication networks. In \cite{R11}, the authors generally illustrated the modeling of NOMA-aided UAV communication networks. In \cite{R12}, NOMA technology was exploited in cooperative uplink transmissions of the static UAV aerial user. For downlink NOMA transmissions between UAV base station (UAV-BS) and multiple ground users, the outage performance was analyzed in \cite{R13,R32} with static UAV-BS, while the joint trajectory design and resource allocation of dynamic UAV-BS and the scheduling design of users were optimized in \cite{R14,R33,R34}.

Due to the openness of wireless environment and the generally increasing security demand of users, wireless secure communications are of utmost concern. In particular, since A2G wiretap channels are potentially with better channel qualities due to LoS conditions, and the movements of UAVs are easily exposed to the surveillance of malicious eavesdroppers, the confidentiality of UAV wireless communications is more challenging to be well-protected. Traditionally, the high-level encryption algorithms are exploited by using shared secret keys, which are not suitable enough for UAV secure communications due to the challenging key management and distribution, significant processing delay, and the vulnerability to strong computation capability \cite{R3}. As a result, physical-layer security (PLS) has been proposed as an important complementary technique for secure wireless communications \cite{RPLS,RPLS2}, especially for 5G application scenarios \cite{RPLS1}, which is key-less and thus promising for UAV secure communications to overcome the aforementioned drawbacks. As summarized in [13], there have recently been several researches on UAV-involved secure communications according to different roles of UAVs. Typically, by acting as aerial base stations, the joint design of trajectory and transmit power of UAV-BS was firstly proposed in [17] to maximize the average secrecy rate performance of the single-user aerial-to-ground transmission link. Then, due to the practical limit on the perfect knowledge of the locations of external eavesdroppers, the authors in [18] investigated the robust design of trajectory and transmit power of UAV-BS to maximize the worst-case secrecy rate. Extended to the multiuser scenario, the authors in [19] considered the joint design of trajectory and transmit power of UAV-BS, as well as the user association policy, to ensure the secrecy fairness among users.

Except for acting as aerial base stations to provide temporary transmission services, UAVs can also be exploited for mobile relaying or friendly jamming to achieve secure cooperation for cellular communications. The secure design of UAV-enabled mobile relaying was firstly investigated in [20] in which the transmit power was optimized under the given trajectory of the UAV-relay. Then, the authors in [21] additionally considered the joint design of transmit power and trajectory of UAV-relay to maximize the achievable secrecy rate performance. For the UAV-enabled friendly jamming, the authors in [22] maximized the average secrecy rate performance of the single-user cellular transmissions by jointly designing the trajectory and jamming power of the UAV-jammer. Additionally, to reflect the impact of imperfect estimated locations of eavesdroppers in the practical scenarios as [18], the robust design of jamming power and UAV-jammer 3D deployment was also proposed in [23].

Under the scenario where multiple UAVs are available, the secrecy outage probability and ergodic secrecy capacity performance were analyzed in [24] and [25] with multiple randomly-located UAV-BSs by using stochastic geometry. However, due to the controllable mobility characteristic of UAVs, it is not reasonable to assume that the locations of UAVs are randomly distributed. As a result, the authors in [26]--[29] investigated the UAV-enabled hybrid secure communications where UAV-enabled friendly jamming was exploited to assist the secure A2G communications enabled by temporary UAV-BSs for secrecy enhancement. It is noted that the inherent ideas to jointly design the resource allocation and trajectory of dual UAVs in the above references are quite similar, while different optimization objectives were considered and different efficient methods to solve the optimization problem were proposed.

Though there have been several researches on physical-layer secure design for NOMA transmissions since the first work \cite{RNOMAPLS}, it is noted that the related security topic on UAV-involved NOMA transmission schemes has not been paid enough attention at this time. Recently, the secrecy performance analysis and optimization of downlink NOMA transmissions between UAV-BS and ground users were investigated in [31]--[34]. However, the secure approaches proposed in [31]--[32] were designed based on the equipped highly-directional multi-antenna in the considered mmWave network, which is generally not practical for small-scale UAV-BS due to its limited payload capacity. Moreover, the authors in [33]--[34] investigated the precoding/beamforming and/or power allocation design in static UAV-BS enabled scenarios, while the inherent mobility of the UAV-BS and the resulted significant DoFs are not exploited in all of the above researches to provide better secrecy performance under multi-user scenarios.

Based on the above discussions, we investigate a downlink secure UAV-BS-enabled NOMA network to serve multiple ground users with distinguished service requirements \cite{R21}, where users are specifically categorized as security-required users (SUs) and quality-of-service (QoS) required users (QUs). In this scenario, it is noted that the data confidentiality of SUs are significantly affected not only by external malicious eavesdroppers as in \cite{R21}, but also by QUs acting as \emph{internal potential eavesdroppers}, which are ``curious'' about the secrecy transmissions and can easily obtain the transmission parameters and codebooks to demodulate and decode the wiretapped symbols. However, the QoS requirements are still regarded as their primary object, which is different from external eavesdroppers. As a result, our goal is to maximize the achievable secrecy performance of SUs, while simultaneously satisfying the QoS requirements of QUs. With respect to this scenario, the time-slotted non-orthogonal transmission scheme is proposed where the flight period of UAV-BS is divided into multiple time slots and a SU and a QU are formed into a user pair in each time slot. This user-pairing strategy can not only satisfy the service requirements of QUs, but also lead to the improved security performance compared with traditional orthogonal transmissions by introducing interference in the superimposed signal against potential eavesdroppers. Furthermore, the reduced complexity of transmission design can be achieved compared with general multi-user non-orthogonal transmissions. In addition, the trajectory design of UAV-BS is of great importance. By proper trajectory design, the coverage area of UAV-BS can be enlarged to provide better service. Moreover, the channel quality of different users can be dynamically changed, which is helpful to enhance the superiority of the main channel to wiretap channels, and thus beneficial to improve the secrecy performance. It should be pointed out that the trajectory design affected by internal potential eavesdroppers in our work is more challenging than those in \cite{R4,R23,R24,R19,R27,R28,R29}, since UAV-BS cannot be just as far away from potential eavesdroppers as possible, and thus needs to be carefully designed to achieve a trade-off between satisfying QoS requirements and improving secrecy performance. In addition, the user-scheduling strategy along with power allocation are also jointly designed. Above all, the main contributions of our paper can be summarized as follows:

\begin{figure}[tbp]
\centering
\includegraphics[width=8cm,height=3.5cm]{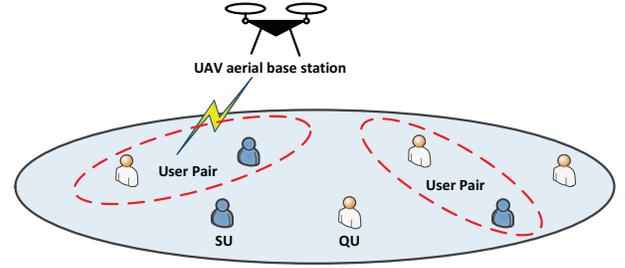}
\caption{The considered UAV-enabled downlink multi-user network. Ground users are categorized by their distinct service requirements. In addition, users are paired for non-orthogonal transmissions.}\label{f_System_model}
\end{figure}

\begin{enumerate}[1)]
\item Subject to the QoS requirements and the constraints on UAV mobility, we jointly optimize the power allocation and trajectory of UAV-BS and the scheduling of ground users to maximize the minimum achievable secrecy rate among SUs, potentially eavesdropped by QUs.

\item For the formulated optimization problem, the penalty-based iterative algorithm is exploited to handle the binary integer scheduling-related variables and to deal with the non-convexity of the corresponding constraints. Then, an efficient iterative algorithm combined with alternative optimization (AO) and successive convex approximation (SCA) methods is proposed.

\item To reduce the computational complexity and obtain a reasonable initial solution for the iterative algorithm, a SU-oriented low-complexity solution is proposed based on the higher priority of secrecy requirements of SUs, the effect of different user pairing methods, and the typical hover-and-fly characteristic of the optimal UAV-BS trajectory.

\end{enumerate}

The rest of this paper is organized as follows: In Section \uppercase\expandafter{\romannumeral2}, we present the system model and the problem formulation. In Section \uppercase\expandafter{\romannumeral3}, the iterative algorithm is proposed. Then, we focus on the low-complexity solution of the problem in Section \uppercase\expandafter{\romannumeral4}. Simulation results are presented in Section \uppercase\expandafter{\romannumeral5} before the conclusion is drawn in Section \uppercase\expandafter{\romannumeral6}.

\section{System Model and Problem Formulation}

\subsection{System Model}
The considered UAV-enabled downlink multi-user network is shown in Fig. 1. With their rapid development, small-scale UAVs are promising to act as aerial platforms to provide temporary wireless communications for the scenario where the traditional cellular network architecture has not been established or has been destroyed due to the natural disasters. In future wireless communications, it has been widely recognized that the service requirements of users will be more and more distinguished due to different kinds of emerging applications. According to the distinct service requirements of different kinds of users, the ground users are categorized as $K$ SUs with high-level secrecy requirements and $M$ QUs with only QoS requirements. Here we make an assumption that the number of SUs in our considered system is no more than that of QUs, which is reasonable in the practical scenarios since it is a general demand to satisfy the basic rate requirements of mobile users and the physical-layer secrecy transmissions can be regarded as their optional value-added services. Without loss of generality, the $k$-th SU and $m$-th QU are located at ${{\mathbf{w}}_k} \in {\mathbb{R}^{2 \times 1}}$ and ${{\mathbf{v}}_m} \in {\mathbb{R}^{2 \times 1}}$, respectively, in the horizontal plane. Each node in the network is equipped with single antenna. It should be pointed out that the above scenario commonly exists in many practical applications. For example, users with high secrecy priority (government officers, etc) will buy the additional physical-layer security services from operators while some others may not. Another typical example is IoT applications such as the Internet of vehicles, where some sensors require confidential data (e.g., states of vehicle engines) while the data for some other sensors may have low or no secrecy requirements (e.g., temperature).

\subsection{Time-Slotted Non-Orthogonal Transmission}
In this section, the time-slotted non-orthogonal transmission scheme is introduced based on different requirements of users, where the flight period of UAV-BS is divided into multiple time slots and the power-domain NOMA transmission scheme is adopted in each time slot to effectively improve the spectral efficiency of the network and provide more opportunities for simultaneous connectivity of ground users. Due to the interference-limited characteristic of NOMA transmissions and the limited computation capacity of transceivers, the user pairing strategy is adopted in our proposed scheme. Specifically, a SU and a QU are paired in each slot in our proposed time-slotted NOMA transmission scheme. At this time, each time slot can be exploited for the transmissions of both SUs and QUs, in order to satisfy the requirements of QUs and simultaneously improve the secrecy performance of SUs. Moreover, the expected signal of the scheduled QU can be exploited as the equivalent jamming signal in our proposed user pairing scheme, which can reduce the eavesdropping quality of the unpaired QUs and furthermore improve the achievable secrecy performance of SUs.

To enlarge the coverage area for providing better service quality, the mobility of UAV-BS should be fully exploited and thus the trajectory design of UAV-BS is of great importance in our proposed scheme. Due to the limited battery capacity, the flight duration of UAV-BS is set as $T$, and the 3D position of UAV-BS is represented as $\left\{ {\left. {{{\left( {{\mathbf{q}}_t^T,H} \right)}} \in {\mathbb{R}^{1 \times 3}}} \right|0 \leqslant t \leqslant T} \right\}$, where ${{\mathbf{q}}_t} \in {\mathbb{R}^{2 \times 1}}$ is the horizontal position of UAV-BS at time instant $t$ and $H$ is the flight height of UAV-BS. It is pointed out that $H$ is set as the constant minimum height to avoid collision with obstacles and enhance the A2G channel quality in the meantime. To facilitate the trajectory design, the flight duration of UAV-BS is discretized into $N$ time slots, where $T = N{\delta _t}$ and ${\delta _t}$ represents the fixed length of transmission time slot balancing the complexity and approximation accuracy, in which the position of UAV-BS is approximately assumed invariant. Then, the horizontal position of UAV-BS at time slot $n$ is denoted as ${\mathbf{q}}\left[ n \right] \in {\mathbb{R}^{2 \times 1}}$. Denote the maximum speed of UAV-BS as ${V_{\max }}$, and the maximum flight distance in each time slot is ${D_{\max }} = {V_{\max }}{\delta _t}$. Then, the mobility constraints are expressed as follows:
\begin{equation}\label{e1}
  {\left\| {{\bf{q}}\left[ {n + 1} \right] - {\bf{q}}\left[ n \right]} \right\|_2} \le {D_{\max }},\;\;n \in {\text{N}} \cup \left\{ 0 \right\},
\end{equation}
where ${\bf{q}}\left[ 0 \right] = {{\bf{q}}_{{\rm{Initial}}}}$ and ${\bf{q}}\left[ {N + 1} \right] = {{\bf{q}}_{{\rm{Final}}}}$ are the pre-determined initial and final horizontal positions of UAV-BS, respectively, and ${\text{N}} \triangleq \left\{ {\left. {n \in \mathbb{Z}\;} \right|1 \leqslant n \leqslant N} \right\}$. For the user-pairing strategy in each time slot, the scheduling constraints are formulated as follows during the flight period:
\begin{equation}\label{e2}
  {a_k}\left[ n \right] \in \left\{ {0,1} \right\}\;,\;{b_m}\left[ n \right] \in \left\{ {0,1} \right\},\;\;\forall n \in {\bf{{\text N}}},
\end{equation}
\begin{equation}\label{e3}
  \sum\nolimits_{k = 1}^K {{a_k}\left[ n \right]}  \le 1\;,\;\sum\nolimits_{m = 1}^M {{b_m}\left[ n \right]}  \le 1,\;\;\forall n \in {\bf{{\text N}}},
\end{equation}
where ${a_k}\left[ n \right]$ and ${b_m}\left[ n \right]$ are scheduling-related binary indicator variables. Specifically, ${a_k}\left[ n \right] = 1$ and ${b_m}\left[ n \right] = 1$ indicate that the $k$-th SU and $m$-th QU are scheduled in the $n$-th time slot.

\subsection{Transmission and Eavesdropping Model}
The transmitted superimposed signal in $n$-th time slot is expressed as
\begin{equation}\label{e4}
  s\left[ n \right] = \sqrt {{\alpha _1}\left[ n \right]{P_{tot}}} {s_1}\left[ n \right] + \sqrt {{\alpha _2}\left[ n \right]{P_{tot}}} {s_2}\left[ n \right],
\end{equation}
where ${{P_{tot}}}$ is the transmit power of UAV-BS, ${s_1}\left[ n \right]$ and ${s_2}\left[ n \right]$ are the expected signals of the scheduled SU and QU, and ${\alpha _1}\left[ n \right]$ and ${\alpha _2}\left[ n \right]$ are the corresponding power allocation coefficients that satisfy
\begin{equation}\label{e5}
  {\alpha _1}\left[ n \right] + {\alpha _2}\left[ n \right] = 1,\;\;{\alpha _1}\left[ n \right] \ge 0,\;\;{\alpha _2}\left[ n \right] \ge 0,\;\;\forall n \in {\bf{{\text N}}}.
\end{equation}
The A2G channels between UAV-BS and ground users are assumed dominated by LoS conditions and thus follow the free-space path loss model \cite{R19,R27,R28,R29}, which is also justified by the 3GPP field measurements in 3GPP TR 36.777 for rural and sub-urban scenarios with the requirement of a certain minimum height of UAVs. Specifically, the channel power gain from UAV-BS to the $k$-th SU and the $m$-th QU in the $n$-th time slot are expressed as
\begin{subequations}\label{e7}
\begin{align}
&{h_k}\left[ n \right] = \frac{{{\beta _0}}}{{{{\left\| {{\bf{q}}\left[ n \right] - {{\bf{w}}_k}} \right\|}^2} + {H^2}}},\\
&{g_m}\left[ n \right] = \frac{{{\beta _0}}}{{{{\left\| {{\bf{q}}\left[ n \right] - {{\bf{v}}_m}} \right\|}^2} + {H^2}}},
\end{align}
\end{subequations}
respectively, where ${{\beta _0}}$ denotes the channel power gain at the reference distance. Then, for the scheduled $k$-th SU in the $n$-th time slot, the received signal is formulated as
\begin{equation}\label{e9}
  {y_k}\left[ n \right] = {h_k}\left[ n \right]\left( {\sqrt {{\alpha _1}\left[ n \right]{P_{tot}}} {s_1}\left[ n \right] \!+ \!\sqrt {{\alpha _2}\left[ n \right]{P_{tot}}} {s_2}\left[ n \right]} \right) + \varepsilon,
\end{equation}
where $\varepsilon \!\sim\! {\mathop{\rm CN}\nolimits} \left( {0,{\sigma ^2}} \right)$ denotes the received noise with variance ${\sigma ^2}$. Due to the higher priority of secrecy requirements of SUs, the SIC process is adopted at the $k$-SU to firstly remove the interference signal with the signal-to-interference-and-noise ratio (SINR)
\begin{equation}\label{e10}
  SIN{R_{k \to m}}\left[ n \right] = \frac{{{h_k}\left[ n \right]{\alpha _2}\left[ n \right]\rho }}{{{h_k}\left[ n \right]{\alpha _1}\left[ n \right]\rho  + 1}},
\end{equation}
where $\rho  \buildrel \Delta \over = {{{P_{tot}}} \mathord{\left/
 {\vphantom {{{P_{tot}}} {{\sigma ^2}}}} \right.
 \kern-\nulldelimiterspace} {{\sigma ^2}}}$ is the transmit signal-to-noise ratio (SNR). After successful SIC process, the received SNR of the $k$-th SU in the $n$-th time slot is denoted as
\begin{equation}\label{e11}
  SNR_k\left[ n \right] = {h_k}\left[ n \right]{\alpha _1}\left[ n \right]\rho .
\end{equation}
Similarly, the received SINR of the expected signal of the $m$-th QU and the resulted eavesdropping SNR with interference cancellation in the $n$-th time slot are formulated as
\begin{subequations}\label{e13}
\begin{align}
&SIN{R_m}\left[ n \right] = \frac{{{g_m}\left[ n \right]{\alpha _2}\left[ n \right]\rho }}{{{g_m}\left[ n \right]{\alpha _1}\left[ n \right]\rho  + 1}},\\
&SNR_{m \to k}^{Eve}\left[ n \right] = {g_m}\left[ n \right]{\alpha _1}\left[ n \right]\rho,
\end{align}
\end{subequations}
respectively. It should be pointed out that the scheduled SU and the unscheduled QUs in our considered system have different prior knowledge about the dynamically-adjusted precoding and codeword set. Without the ability to obtain the prior information in the transmission slot, the unscheduled QU cannot remove the interference for successful SIC process. Moreover, for the potential internal eavesdroppers, it is reasonable to assume that the unscheduled QUs will not adopt the complex signal processing technologies in advance, but directly regard the expected signal of the secrecy transmission of the scheduled SU as the target signal, and then decode the target signal under the impact of the interference and background noise. As a result, the eavesdropping SINR of unscheduled QUs in the $n$-th time slot is expressed as
\begin{equation}\label{e15}
  SINR_{\tilde m \to k}^{Eve}\left[ n \right] = \frac{{{g_{\tilde m}}\left[ n \right]{\alpha _1}\left[ n \right]\rho }}{{{g_{\tilde m}}\left[ n \right]{\alpha _2}\left[ n \right]\rho  + 1}}\;,\;\;\forall \tilde m \ne m.
\end{equation}
Then, the achievable secrecy rate of the $k$-th SU in the $n$-th time slot is denoted as
\begin{equation}\label{e16}
  R_k^{\left( s \right)}\left[ n \right] \buildrel \Delta \over = {\left\{ {{\log _2}\left( {1 + SN{R_k}\left[ n \right]} \right) - {R_{Eve}}\left[ n \right]} \right\}^ + },
\end{equation}
where ${\left\{ a \right\}^ + } \!\buildrel \Delta \over = \!\max \left( {a,0} \right)$ and ${R_{Eve}}\left[ n \right]$ denotes the maximum eavesdropping rate in the $n$-th slot, which is provided at the top of the next page.
\begin{figure*}[ht]
\begin{equation}\label{e17}
  {R_{Eve}}\left[ n \right]\! \buildrel \Delta \over =\!\!\! \mathop {\max }\limits_{m = 1,2,...,M} \left\{ {{b_m}\left[ n \right]{{\log }_2}\left( {1\! + \!SNR_{m \to k}^{Eve}\left[ n \right]} \right),\left( {1 \!-\! {b_m}\left[ n \right]} \right){{\log }_2}\left( {1 \!+ \!SINR_{m \to k}^{Eve}\left[ n \right]} \right)} \right\}
\end{equation}
\end{figure*}

\subsection{Problem Formulation}
We jointly design the optimal power allocation strategy, user scheduling and UAV-BS trajectory to maximize the minimum achievable secrecy rate among SUs as follows:
\begin{subequations}\label{P1}
\begin{align}
&\mathop {\max }\limits_{{\mathbf{P}},{\bf{A}},{\bf{Q}}} \;\mathop {\min }\limits_{k = 1,2,...,K} \;\frac{1}{N}\sum\limits_{n = 1}^N {{a_k}\left[ n \right]R_k^{\left( s \right)}\left[ n \right]} \label{P1e1}\\
&\;\;\;{\rm{s}}{\rm{.t}}{\rm{.}}\;\;\;\sum\limits_{n = 1}^N {{b_m}\left[ n \right]R_m^{\left( {QoS} \right)}\left[ n \right]}  \ge {\gamma _m}\;,\;m = 1,2,...,M \label{P1e2} \\
&\;\;\;\;\;\;\;\;\;\;\;(\textrm{\ref{e1}}) \ , \ (\textrm{\ref{e2}}) \ , \ (\textrm{\ref{e3}}) \ \textrm{and} \ (\textrm{\ref{e5}}), \label{P1e3}
\end{align}
\end{subequations}
where $ {\mathbf{P}} \triangleq \left\{ {\left. {{\alpha _i}\left[ n \right] \in \mathbb{R}\;} \right|i \in \left\{ {1,2} \right\},\forall n} \right\}$ represents the set of power allocation coefficients, ${\bf{A}} \buildrel \Delta \over = \left\{ {\left. {{a_k}\left[ n \right],{b_m}\left[ n \right]\;} \right|\forall k,m,n} \right\}$ denotes the set of scheduling indicators during the flight period, ${\bf{Q}} \buildrel \Delta \over = \left\{ {\left. {{\bf{q}}\left[ n \right] \in {\mathbb{R}^{2 \times 1}}} \right|n \in {\text{N}}} \right\}$ is the set of horizontal positions of UAV-BS, $R_m^{\left( {QoS} \right)}\left[ n \right] \buildrel \Delta \over = {\log _2}\left( {1 + SIN{R_m}\left[ n \right]} \right)$ denotes the rate of the scheduled $m$-th QU in the $n$-th time slot, and ${\gamma _m}$ is the minimum QoS requirement of the $m$-th QU.

Due to the complex objective function and the non-convex constraints (\ref{e2}) and (\ref{e3}) related to binary scheduling indicator variables, the aforementioned problem is difficult to be directly solved. In the following sections, we firstly propose an iteration-based algorithm with SCA methods in an alternative manner. Then, a low-complexity algorithm is proposed by exploiting the typical hover-and-fly characteristic of UAV-BS and the distinct requirements of users.

\section{Iteration-Based Algorithm}
Due to the high complexity to jointly solve problem (\ref{P1}) caused by the coupled optimization variables in the objective functions and constraints, we divide the optimization variables into three parts, consisting of scheduling-related indicators, power allocation coefficients, and
two-dimensional (2D) horizontal positions of UAV-BS during the time-slotted flight period. Then, the non-convex objective function and constraints can be approximately transformed, and the divided sub-problems can be solved in an alternative manner to obtain a sub-optimal solution at this time.

\subsection{User Scheduling Sub-Problem}
Based on the given trajectory and power allocation strategy during the flight period of UAV-BS, the related slack variables are introduced and the user scheduling sub-problem is transformed as
\begin{subequations}\label{P2}
\begin{align}
&\mathop {\max }\limits_{{\bf{A}},{\bf{\Phi }},{\bf{\Theta }},{\bf{\Gamma }},{\tau }} \;\;\;\;\;\tau  \label{P2e1}\\
&\;\;\;\;{\rm{s}}{\rm{.t}}{\rm{.}}\;\;\;\;\frac{1}{N}\sum\limits_{n = 1}^N {{\phi _k}\left[ n \right]}  \ge \tau  \label{P2e2} \\
&\;\;\;\;\;\;\;\;\;\;\;\;\;{a_k}\left[ n \right] \cdot {\theta _k}\left[ n \right] \ge {\phi _k}\left[ n \right],\;\forall k,n  \label{P2e3} \\
&\;\;\;\;\;\;\;\;\;\;\;\;\;{\log _2}\left( {1 + SN{R_k}\left[ n \right]} \right) \!-\! {\varsigma _k}\left[ n \right] \ge {\theta _k}\left[ n \right],\;\forall k,n  \label{P2e4} \\
&\;\;\;\;\;\;\;\;\;\;\;\;\;{b_m}\left[ n \right] \cdot {\log _2}\left( {1 + SNR_{m \to k}^{Eve}\left[ n \right]} \right) \le {\varsigma _k}\left[ n \right] \label{P2e5} \\
&\;\;\;\;\;\;\;\;\;\;\;\;\left( {1\! -\! {b_m}\left[ n \right]} \right) \!\cdot\! {\log _2}\left( {1 \!+\! SINR_{m \to k}^{Eve}\left[ n \right]} \right) \!\le\! {\varsigma _k}\left[ n \right]  \label{P2e6} \\
&\;\;\;\;\;\;\;\;\;\;\;\;\sum\limits_{n = 1}^N {{b_m}\left[ n \right] \cdot {{\log }_2}\left( {1 + SIN{R_m}\left[ n \right]} \right)}  \ge {\gamma _m} \label{P2e7}\\
&\;\;\;\;\;\;\;\;\;\;\;\;(\textrm{\ref{e2}}) \ \textrm{and} \ (\textrm{\ref{e3}})
\end{align}
\end{subequations}
where ${\bf{\Phi }} \buildrel \Delta \over = \left\{ {\left. {{\phi _k}\left[ n \right] \in {\mathbb{R}}} \right|\forall k,n} \right\}$, ${\bf{\Theta }} \buildrel \Delta \over = \left\{ {\left. {{\theta _k}\left[ n \right] \in {\mathbb{R}}} \right|\forall k,n} \right\}$, and ${\bf{\Gamma }} \buildrel \Delta \over = \left\{ {\left. {{\varsigma _k}\left[ n \right] \in {\mathbb{R}}} \right|k \in \left\{ {1,2,...,K} \right\},\forall n} \right\}$ are the introduced slack variable sets. The difficulty of solving the above transformed optimization problem lies in the non-convexity of constraint (\ref{P2e3}) and the existence of binary integer variables. Therefore, we firstly introduce continuous variables ${\bf{\tilde A}} \buildrel \Delta \over = \left\{ {\left. {{{\tilde a}_k}\left[ n \right] \in {\mathbb{R}},{{\tilde b}_m}\left[ n \right] \in {\mathbb{R}}} \right|\forall k,m,n} \right\}$ and relax the original scheduling-related optimization variables into continuous ones at the same time. In such a context, additional equality constraints are introduced that
\begin{equation}\label{e18}
  {a_k}\left[ n \right]\left( {1 - {{\tilde a}_k}\left[ n \right]} \right) = 0,\;{a_k}\left[ n \right] = {{\tilde a}_k}\left[ n \right]
\end{equation}
\begin{equation}\label{e19}
  {b_m}\left[ n \right]\left( {1 - {{\tilde b}_m}\left[ n \right]} \right) = 0,\;{b_m}\left[ n \right] = {{\tilde b}_m}\left[ n \right]
\end{equation}
are satisfied for $\forall k,m,n$ and thus the optimal relaxed continuous variables are restricted to have the binary integer forms. Then, the penalty terms with respect to the above equalities are introduced into the objective function to reflect the impact of constraint relaxation on the optimal solutions\footnote{Though the binary variable relaxation method can be applied for user-scheduling design as in \cite{Rfootnote}, the penalty-based algorithm is more appropriate in this paper without the reconstruction of binary integer scheduling variables, due to the requirement of the fixed length of each transmission time slot.}. The resulted optimization problem is formulated as follows:
\begin{subequations}\label{P3}
\begin{align}
&\begin{array}{l}
\mathop {\max }\limits_{{\bf{A}},{\bf{\Phi }},{\bf{\Theta }},{\bf{\Gamma }},{\bf{\tilde A}}}\! \tau \! -\! \eta {\sum\limits_{k,n} {\left( {{{\left( {{a_k}\left[ n \right] \!-\! {{\tilde a}_k}\left[ n \right]} \right)}^2} \!+\! {{\left( {{a_k}\left[ n \right]\left( {1 \!-\! {{\tilde a}_k}\left[ n \right]} \right)} \right)}^2}} \right)} } \\
\;\;\;\;\;\;\;\;\; - \eta {\sum\limits_{m,n} {\left( {{{\left( {{b_m}\left[ n \right] \!-\! {{\tilde b}_m}\left[ n \right]} \right)}^2} \!+\! {{\left( {{b_m}\left[ n \right]\left( {1\! -\! {{\tilde b}_m}\left[ n \right]} \right)} \right)}^2}} \right)} }
\end{array}  \label{P3e1}\\
&\;\;\;\;\;\;{\rm{s}}{\rm{.t}}{\rm{.}}\;\;\;\;\;\;\;0 \le {a_k}\left[ n \right] \le 1,\;\;\;0 \le {b_m}\left[ n \right] \le 1,\;\forall k,m,n\;  \label{P3e2} \\
&\;\;\;\;\;\;\;\;\;\;\;\;\;\;\;\;\;(\textrm{\ref{P2e2}}) - (\textrm{\ref{P2e7}}) \ \textrm{and} \ (\textrm{\ref{e3}}), \label{P3e4}
\end{align}
\end{subequations}
where $\eta $ is denoted as the penalty coefficient, and constraints (\ref{P3e2}) is introduced to accelerate the rate of convergence. Based on the above problem, an iteration-based algorithm is proposed to update $\eta $ through iterations. The values of penalty terms in the objective function are finally under a predefined threshold when the algorithm converges, which means that the optimal solutions of the relaxed variables are satisfied to approximately have the binary integer forms.

In each iteration, it is observed that the introduced variables ${{\bf{\tilde A}}}$ only exist in the objective function of problem (\ref{P3}). As a result, the optimization variables can be equivalently divided into ${{\bf{\tilde A}}}$ and $\left\{ {{\bf{A}},{\bf{\Phi }},{\bf{\Theta }},{\bf{\Gamma }}} \right\}$, and the optimization problem can be iteratively solved in an alternative manner. In the $i$-th inner iteration, with the fixed ${{{\bf{A}}^{\left( {i - 1} \right)}}}$ obtained in the $(i-1)$-th iteration, the optimal ${{{{\bf{\tilde A}}}^{\left( i \right)}}}$ can be derived via the first-order derivative of the objective function as follows:
\begin{equation}\label{e20}
  \tilde a_k^{\left( i \right)}\left[ n \right] = {{a_k^{\left( {i \!-\! 1} \right)}\left[ n \right]\left( {1\! +\! a_k^{\left( {i \!-\! 1} \right)}\left[ n \right]} \right)} \mathord{\left/
 {\vphantom {{a_k^{\left( {i \!-\! 1} \right)}\left[ n \right]\left( {1 \!+ \!a_k^{\left( {i \!-\! 1} \right)}\left[ n \right]} \right)} {1 + {{\left( {a_k^{\left( {i \!-\! 1} \right)}\left[ n \right]} \right)}^2}}}} \right.
 \kern-\nulldelimiterspace} {1 + {{\left( {a_k^{\left( {i \!-\! 1} \right)}\left[ n \right]} \right)}^2}}},
\end{equation}
\begin{equation}\label{e21}
  \tilde b_m^{\left( i \right)}\left[ n \right] = {{b_m^{\left( {i \!-\! 1} \right)}\left[ n \right]\left( {1 + b_m^{\left( {i \!-\! 1} \right)}\left[ n \right]} \right)} \mathord{\left/
 {\vphantom {{b_m^{\left( {i\! -\! 1} \right)}\left[ n \right]\left( {1 + b_m^{\left( {i \!-\! 1} \right)}\left[ n \right]} \right)} {1 + {{\left( {b_m^{\left( {i\! -\! 1} \right)}\left[ n \right]} \right)}^2}}}} \right.
 \kern-\nulldelimiterspace} {1 + {{\left( {b_m^{\left( {i\! - \!1} \right)}\left[ n \right]} \right)}^2}}}.
\end{equation}

\begin{algorithm}[t]
\setstretch{1}
\caption{Penalty-Based Iterative Algorithm for User-Scheduling Sub-Problem}\label{A1}
\begin{algorithmic}[1]
\STATE \textbf{Initialize}: Fixed power allocation coefficients ${\bf{\alpha }}$ and UAV-BS trajectory ${\bf{Q}}$;
\STATE \textbf{Set}:  Outer iteration index as $i = 0$, inner iteration index as $j = 0$, penalty coefficient $\eta  = 1$, increasing factor $c = 2$, maximum iteration step $L = 20$, and the threshold $\omega  = {10^{ - 3}}$;
\STATE \textbf{Repeat (Outer Loop)}: $i = i + 1$;
\STATE \ \ \ \textbf{Initialize}: $\left\{ {{{\bf{A}}^{\left( 0 \right)}},{{\bf{\Phi }}^{\left( 0 \right)}}} \right\}$ and ${R_{old}} = {10^{ - 7}}$;
\STATE \ \ \ \textbf{Repeat (Inner Loop)}: $j = j + 1$;
\STATE \ \ \ \ \ \ Obtain the optimal ${{{{\bf{\tilde A}}}^{\left( j \right)}}}$ by (\ref{e20}) and (\ref{e21}) with fixed ${{{\bf{A}}^{\left( {j - 1} \right)}}}$;
\STATE \ \ \ \ \ \ Solve problem (\ref{P3}) and obtain the optimal ${{{\bf{A}}^{\left( {j} \right)}}}$ and objective function value $R$;
\STATE \ \ \ \textbf{Until}: ${{\left( {R - {R_{old}}} \right)} \mathord{\left/
 {\vphantom {{\left( {R - {R_{old}}} \right)} {{R_{old}}}}} \right.
 \kern-\nulldelimiterspace} {{R_{old}}}} \le \omega $ or $j > L$; \ \ \textbf{Else}: ${{R_{old}} = R}$;
\STATE \textbf{Calculate}: $\kappa $ representing the maximum value of the penalty terms;
\STATE \textbf{Until}: $\kappa  \le \omega $ or $i > L$; \ \ \textbf{Else}: $\eta  = c\eta $;
\end{algorithmic}
\end{algorithm}

\noindent{Based on the obtained ${{{{\bf{\tilde A}}}^{\left( i \right)}}}$, we aim to solve problem (\ref{P3}) to optimize the other variables. Due to the coupling of optimization variables in constraint (\ref{P2e3}), we exploit the first-order Taylor approximation method and thus the constraint (\ref{P2e3}) in $i$-th inner iteration is transformed as}
\begin{subequations}\label{e22}
\begin{align}
& \;\;\;\;\;\;a_k^{\left( i \right)}\left[ n \right] \cdot \theta _k^{\left( i \right)}\left[ n \right] \ge \phi _k^{\left( i \right)}\left[ n \right] \notag \\
& \Rightarrow \frac{{{{\left( {a_k^{\left( i \right)}\left[ n \right] \!+\! \theta _k^{\left( i \right)}\left[ n \right]} \right)}^2}}}{4} \!-\! \frac{{{{\left( {a_k^{\left( i \right)}\left[ n \right] \!-\! \theta _k^{\left( i \right)}\left[ n \right]} \right)}^2}}}{4} \ge \phi _k^{\left( i \right)}\left[ n \right] \tag{21} \\
&  \Rightarrow \frac{{{{\left( {a_k^o\left[ n \right] + \theta _k^o\left[ n \right]} \right)}^2}}}{4} + \frac{{\left( {a_k^o\left[ n \right] + \theta _k^o\left[ n \right]} \right)\left( {{a_k}\left[ n \right] + {\theta _k}\left[ n \right]} \right)}}{2}\notag \\
& \;\;\;\;\;- \frac{{{{\left( {a_k^{\left( i \right)}\left[ n \right] - \theta _k^{\left( i \right)}\left[ n \right]} \right)}^2}}}{4} \ge \phi _k^{\left( i \right)}\left[ n \right], \notag
\end{align}
\end{subequations}
where $a_k^o\left[ n \right] = a_k^{\left( {i - 1} \right)}\left[ n \right]$ and $\theta _k^o\left[ n \right] = \theta _k^{\left( {i - 1} \right)}\left[ n \right]$ are obtained in the $(i-1)$-th iteration, representing the fixed points of first-order approximation. After substituting constraint (\ref{e22}), the transformed sub-problem satisfies the requirements of convex optimization, and thus can be efficiently solved. The proposed penalty-based algorithm is summarized as \emph{Algorithm \ref{A1}}.

\begin{remark}
It is mentioned that the scheduled SU should have a better channel condition than the paired QU in each slot to ensure the perfect SIC process, which may be required as an additional constraint for optimization. However, since the scheduled QU has the ability to decode, reconstruct and remove the interference signal for interception, if the channel quality from UAV-BS to the scheduled QU is better than that to the scheduled SU, the eavesdropping channel capacity is superior to the main channel capacity, which will lead to the zero achievable secrecy capacity. As a result, for the user-scheduling design, it is obvious that the channel quality of the scheduled QU will not be better than that of the scheduled SU to maximize the achievable secrecy rate performance, and thus the perfect SIC process can be reasonably satisfied.
\end{remark}

\subsection{Power Allocation Sub-Problem}
Based on the given UAV-BS horizontal trajectory and user scheduling policy, the slack variables are introduced and the power allocation sub-problem is formulated as
\begin{subequations}\label{P4}
\begin{align}
&\mathop {\max }\limits_{{\mathbf{P}},{\mathbf{u}},\tau } \;\;\;\;\;\tau   \label{P4e1}\\
&\;\;{\rm{s}}.{\rm{t}}.\;\;\;\;\;\frac{1}{N}\sum\limits_{n \in {{\bf{\Omega }}_k}} {\left( {{{\log }_2}\left( {1 \!+\! \rho {h_k}\left[ n \right]{\alpha _1}\left[ n \right]} \right) \!-\! \mu \left[ n \right]} \right)}  \ge \tau  \label{P4e2} \\
&\;\;\;\;\;\;\;\;\;\;\;\;{\log _2}\left( {1 + \rho {g_{{m^*}\left[ n \right]}}\left[ n \right]{\alpha _1}\left[ n \right]} \right) \le \mu \left[ n \right],  \label{P4e3} \\
&\;\;\;\;\;\;\;\;\;\;\;\;{\log _2}\left( {1\! + \!\frac{{\rho {g_m}\left[ n \right]{\alpha _1}\left[ n \right]}}{{\rho {g_m}\left[ n \right]{\alpha _2}\left[ n \right]\! +\! 1}}} \right)\! \le\! \mu \left[ n \right],\;\forall m \!\ne\! {m^*}\left[ n \right] \label{P4e4} \\
&\;\;\;\;\;\;\;\;\;\;\;\sum\limits_{n \in {{\bf{\Psi }}_m}} {{{\log }_2}\left( {1\! +\! \frac{{\rho {g_m}\left[ n \right]{\alpha _2}\left[ n \right]}}{{\rho {g_m}\left[ n \right]{\alpha _1}\left[ n \right] \!+\! 1}}} \right)} \! \ge\! {\gamma _m}, \textrm{and} \;(\textrm{\ref{e5}}) \label{P4e5}
\end{align}
\end{subequations}
where ${\mathbf{u}} \buildrel \Delta \over = \left\{ {\left. {\mu \left[ n \right]\;} \in {\mathbb{R}} \right|n \in {\text{N}}} \right\}$ is the introduced slack variable set, ${{\bf{\Omega }}_k} \buildrel \Delta \over = \left\{ {\left. {n \in {\text{N}}} \right|{a_k}\left[ n \right] = 1} \right\}$ and ${{\bf{\Psi }}_m} \buildrel \Delta \over = \left\{ {\left. {n \in {\text{N}}} \right|{b_m}\left[ n \right] = 1} \right\}$ represent the collection of scheduled time slots of the $k$-th SU and $m$-th QU, respectively, and ${m^*}\left[ n \right] \buildrel \Delta \over = \left\{ {\left. {m} \right|{b_m}\left[ n \right] = 1} \right\}$ is defined as the index of the scheduled QU in the $n$-th time slot. By exploiting the constraint ${\alpha _1}\left[ n \right] + {\alpha _2}\left[ n \right] = 1, \forall n$, the constraint (\ref{P4e4}) is transformed as
\begin{equation}\label{e23}
  {\log _2}\left( {1 + \rho {g_m}\left[ n \right]} \right) - {\log _2}\left( {1 + \rho {g_m}\left[ n \right]{\alpha _2}\left[ n \right]} \right) \le \mu \left[ n \right],
\end{equation}
which is observed to satisfy the requirements of convex optimization. Then, the first-order Taylor approximation method can be exploited to approximately transform other non-convex constraints as in the previous sub-section, which are omitted in this sub-section due to the page limit.
After substituting the above approximated constraints into problem (\ref{P4}), the transformed sub-problem satisfies the requirements of convex optimization and thus can be efficiently solved.

\subsection{Trajectory Design Sub-Problem}
Based on the given user scheduling policy and the fixed power allocation coefficients, the slack variables are introduced and the trajectory design sub-problem is formulated as
\begin{subequations}\label{P5}
\begin{align}
&\mathop {\max }\limits_{{\bf{Q}},{\mathbf{u}},\tau } \;\;\;\;\;\tau   \label{P5e1}\\
&\;\;{\rm{s}}.{\rm{t}}.\;\;\;\frac{1}{N}\sum\limits_{n \in {{\bf{\Omega }}_k}}\! {\left(\! {{{\log }_2}\!\left(\! {1\! +\! \frac{{\rho {\beta _0}{\alpha _1}\left[ n \right]}}{{{{\left\| {{\bf{q}}\left[ n \right]\! -\! {{\bf{w}}_k}} \right\|}^2} \!+\! {H^2}}}} \right)\! -\! \mu \left[ n \right]} \!\right)} \! \ge\! \tau \label{P5e2} \\
&\;\;\;\;\;\;\;\;\;\;{\log _2}\left( {1 + \frac{{\rho {\beta _0}{\alpha _1}\left[ n \right]}}{{{{\left\| {{\bf{q}}\left[ n \right] - {{\bf{v}}_{{m^*}\left[ n \right]}}} \right\|}^2} + {H^2}}}} \right) \le \mu \left[ n \right], \label{P5e3} \\
&\;\;\;\;\;\;\;\;\;\;{\log _2}\!\left(\! {1 \!+\! \frac{{\rho {\beta _0}}}{{{{\left\| {{\bf{q}}\left[ n \right] \!-\! {{\bf{v}}_m}} \right\|}^2}\! +\! {H^2}}}} \!\right)\notag\\
&\;\;\;\;\;\;\;\!-\! {\log _2}\!\left( \!{1 \!+\! \frac{{\rho {\beta _0}{\alpha _2}\left[ n \right]}}{{{{\left\| {{\bf{q}}\left[ n \right] \!-\! {{\bf{v}}_m}} \right\|}^2}\! +\! {H^2}}}}\! \right) \!\le \mu \left[ n \right] \label{P5e4} ,\forall m \ne {m^*}\left[ n \right] \\
&\;\;\;\;\;\;\;\;\sum\limits_{n \in {{\bf{\Psi }}_m}}\!\! {\left(\! {{{\log }_2}\!\left( {1 \!+\! \frac{{\rho {\beta _0}}}{{{{\left\| {{\bf{q}}\left[ n \right]\! -\! {{\bf{v}}_m}} \right\|}^2}\! +\! {H^2}}}} \!\right) } \!\right)} \label{P5e5}\notag \\
&\;\;\;\;-\sum\limits_{n \in {{\bf{\Psi }}_m}}\!\! {\left(\! {{{\log }_2}\!\left(\! {1\! +\! \frac{{\rho {\beta _0}{\alpha _1}\left[ n \right]}}{{{{\left\| {{\bf{q}}\left[ n \right] - {{\bf{v}}_m}} \right\|}^2}\! +\! {H^2}}}} \!\right)} \!\right)}  \ge {\gamma _m} \\
&\;\;\;\;\;\;\;\;\;{\left\| {{\bf{q}}\left[ {n + 1} \right] - {\bf{q}}\left[ n \right]} \right\|_2} \le {D_{\max }},\;\;\forall n \in {\bf{{\text N}}} \cup \left\{ 0 \right\}. \label{P5e6}
\end{align}
\end{subequations}
To solve problem (\ref{P5}), the slack variables, which denote the lower bound of the squared-distance between UAV-BS and QUs, are firstly introduced as
\begin{equation}\label{e27}
  {\pi _m}\left[ n \right] \le {\left\| {{\bf{q}}\left[ n \right] - {{\bf{v}}_m}} \right\|^2},\;\;\forall m,n.
\end{equation}
As a result, the corresponding term in constraint (\ref{P5e3}) can be approximately transformed as
\begin{equation}\label{e28}
{\log _2}\!\left(\! {1 \!+\!\! \frac{{\rho {\beta _0}{\alpha _1}\left[ n \right]}}{{{{\left\| {{\bf{q}}\left[ n \right] \!-\! {{\bf{v}}_{{m^*}\left[ {\rm{n}} \right]}}} \right\|}^2} \!\!+\!\! {H^2}}}} \!\right)\!\! \le \!{\log _2}\!\left(\! {1\! +\!\! \frac{{\rho {\beta _0}{\alpha _1}\left[ n \right]}}{{{\pi _{{m^*}\left[ {\rm{n}} \right]}}\left[ n \right] \!+\! {H^2}}}} \!\right)
\end{equation}
which is a convex function with respect to ${{\pi _{{m^*}\left[ {\rm{n}} \right]}}\left[ n \right]}$, and thus the non-convexity of constraint (\ref{P5e3}) is handled. To deal with the non-convexity introduced by other constraints, the first-order Taylor approximation method is exploited to transform the non-convex term into its linear form as in the previous sub-sections, which are omitted in this sub-section due to the page limit.
It should be pointed out that the introduced constraint (\ref{e27}) with respect to the lower bound on the square of distance is also non-convex, and thus should be transformed as
\begin{subequations}\label{e35}
\begin{align}
&{\pi _m}\left[ n \right] \!\le\! {\left\| {{{\bf{q}}^{\rm{o}}}\left[ n \right] \!-\! {{\bf{v}}_m}} \right\|^2} \!+\! 2{\left( {{{\bf{q}}^{\rm{o}}}\left[ n \right] \!-\! {{\bf{v}}_m}} \right)^T}\left( {{\bf{q}}\left[ n \right] \!-\! {{\bf{q}}^{\rm{o}}}\left[ n \right]} \right)\notag \\
&\;\;\;\;\;\;\;\;\le {\left\| {{\bf{q}}\left[ n \right] - {{\bf{v}}_m}} \right\|^2}\tag{27}.
\end{align}
\end{subequations}
Based on the aforementioned linear approximations, the transformed optimization sub-problem satisfies the requirements of convex optimization and thus can be efficiently solved.

\subsection{Overall Algorithm and Computational Complexity Analysis}
According to the aforementioned solutions for each sub-problem, the overall AO-based iterative algorithm is summarized as \emph{Algorithm \ref{A2}}. In general, the overall algorithm can converge to a stationary point, which is a sub-optimal solution to problem (\ref{P1}). The detailed proof of the convergence can be found in \cite{R4}, and thus are left out in this work for the brevity.

Then, it is noted that problem (\ref{P3}) in each iteration of \emph{Algorithm \ref{A1}} is a linear programming (LP) and the complexity of solving a LP is $O\left( {n_{LP}^2{m_{LP}}} \right)$, where ${{m_{LP}}}$ denotes the number of constraints and ${{n_{LP}}}$ is the dimension of optimization variables \cite{R35}. Specifically, we have ${m_{LP}} = \left(4K + 2\left( {K + 1} \right)M \right)N$ and ${n_{LP}} = \left( {4K + M} \right)N $. Therefore, the computational complexity of \emph{Algorithm \ref{A1}} is denoted as ${\varpi _1} = O\left( {{\eta_1 }{\eta_2 }\left(4K + 2\left( {K + 1} \right)M \right){\left( {4K + M} \right)}^2{N}^3} \right)$, where ${\eta_1 }$ and ${\eta_2 }$ represent the numbers of inner and outer iterations of \emph{Algorithm \ref{A1}}, respectively. Since the power allocation sub-problem is also a LP, the computational complexity can be similarly represented as ${\varpi _2} = O\left( 4{\left(K + M + \left( {M + 2} \right)N \right){N}^2} \right)$. As for the trajectory design sub-problem, since the transformed constrains are all convex quadratic constraints, the computational complexity is denoted at the top of the next page, as in \cite{R36,R37}. Therefore, the overall computational complexity of \emph{Algorithm \ref{A2}} can be represented as ${\varpi} = O\left({\eta_3 }\left( {{\varpi _1} + {\varpi _2} + {\varpi _3}} \right)\right)$, where ${\eta_3 }$ denotes the number of iterations for \emph{Algorithm \ref{A2}} to reach the convergence.

\begin{figure*}[ht]
\begin{equation}\label{eCom}
  {\varpi _3} = O\left( \sqrt {2{\left(K + M + \left( {M + 1} \right)N + 1 \right)}} \left( \left( M + 3\right)N\right) \left( 9KN^2 + 9MN\left( N+1\right) + 11N - 8\right)\right)
\end{equation}
\end{figure*}

\begin{algorithm}[t]
\setstretch{1}
\caption{AO-Based Iterative Algorithm}\label{A2}
\begin{algorithmic}[1]
\STATE \textbf{Initialize}: User scheduling indicators ${{\bf{A}}^{\left( 0 \right)}}$, power allocation coefficients ${{\mathbf{P}}^{\left( 0 \right)}}$ and UAV-BS trajectory ${{\bf{Q}}^{\left( 0 \right)}}$;
\STATE \textbf{Set}:  Iteration index as $i = 0$ and initial objective value ${R_{old}} = {10^{ - 7}}$;
\STATE \textbf{Set}: The maximum iteration step $L = 20$ and the threshold $\omega  = {10^{ - 3}}$;
\STATE \textbf{Repeat}: $i = i + 1$;
\STATE \ \ \ \textbf{Calculate}: The optimal ${{\bf{A}}^{\left( i \right)}}$ by \emph{Algorithm \ref{A1}} with fixed ${{{\bf{A}}^{\left( {i - 1} \right)}}}$, ${{\mathbf{P}}^{\left( i - 1 \right)}}$, and ${{\bf{Q}}^{\left( i - 1 \right)}}$;
\STATE \ \ \ \textbf{Calculate}: The optimal ${{\mathbf{P}}^{\left( i \right)}}$ by solving problem (\ref{P4}) with fixed ${{{\bf{A}}^{\left( {i} \right)}}}$, ${{\mathbf{P}}^{\left( i - 1 \right)}}$, and ${{\bf{Q}}^{\left( i - 1 \right)}}$;
\STATE \ \ \ \textbf{Calculate}: The optimal ${{\bf{Q}}^{\left( i \right)}}$ by solving problem (\ref{P5}) with fixed ${{{\bf{A}}^{\left( {i} \right)}}}$, ${{\mathbf{P}}^{\left( i \right)}}$, and ${{\bf{Q}}^{\left( i - 1 \right)}}$;
\STATE \ \ \ Obtain the current optimal value of the objective function as $R$;
\STATE \textbf{Until}: ${{\left( {R - {R_{old}}} \right)} \mathord{\left/
 {\vphantom {{\left( {R - {R_{old}}} \right)} {{R_{old}}}}} \right.
 \kern-\nulldelimiterspace} {{R_{old}}}} \le \omega $ or $i > L$; \ \ \textbf{Else}: ${{R_{old}} = R}$;
\end{algorithmic}
\end{algorithm}

\section{Low-Complexity Solution}
Though the alternation-based iterative algorithm to effectively solve problem (\ref{P1}) is proposed in the previous section, it is pointed out that there exist the following drawbacks:
\begin{enumerate}[1)]
\item Due to the iterative processing, especially in the double-loop penalty-based \emph{Algorithm \ref{A1}}, the computational complexity of \emph{Algorithm \ref{A2}} is significantly high to obtain a sub-optimal solution as analyzed in Section \uppercase\expandafter{\romannumeral3}-E.

\item For the initialization step, it is noted that a feasible solution of problem (\ref{P1}) needs to be obtained in advance. It should be demonstrated that the obtained sub-optimal solution through multiple iterations is highly related to the initial solution, and thus we have to particularly choose a reasonable initial solution for \emph{Algorithm \ref{A2}}, which requires a systematic approach.

\end{enumerate}
Based on the above analyses, we prefer to design a low-complexity algorithm to solve problem (\ref{P1}) by exploiting the hover-and-fly characteristic of the generally optimal UAV trajectory, the distinct requirements and priorities of SUs and QUs, and the effect of different user pairing policies in non-orthogonal transmissions in this section. Similar to Section \uppercase\expandafter{\romannumeral3}, the overall problem is decoupled into three aspects, consisting of the user-scheduling design, the UAV-BS trajectory design, and finally the power allocation design.

\subsection{User-Scheduling Design}
There exist the following two candidate schemes for our user-pairing design:
\begin{enumerate}[1)]
\item The scheduled SU is paired with the worst-channel-quality QU. Therefore, there exists a sufficient gap between the channel quality of the scheduled SU and QU, which is useful for NOMA to provide better quality compared with traditional OMA. Moreover, only the worst-channel-quality QU has the ability to remove the interference by SIC, and thus the eavesdropping threat by QUs can be reduced.

\item The scheduled SU is paired with the best-channel-quality QU. In this case, the best capacity region performance can be achieved among all the possible choices of user pairing.

\end{enumerate}

Due to the priority of security requirements of SUs and the relatively low QoS requirements of QUs in our considered scenario, we select the first user-pairing scheme to achieve better secrecy performance. It is noted that due to the fact that the trajectory design is mainly determined by the locations of SUs, the worst-channel-quality QU can thus be equivalently approximated as the most distant QU away from the scheduled SU, which can be mathematically expressed as
     \begin{equation}
 \tilde m = \arg \;\mathop {min}\limits_{m = 1,2,...,M} {d_{k \to m}}\;\;\text{if}\;{a_k} = 1.
 \end{equation}
    However, for ease of facilitation and to reflect the fairness of service for each user during the flight period, based on the assumption that there exist more QUs than SUs in our investigated system, the maximal number of the possible user pairs with each SU is firstly defined as $L = ceil\left( {{M \mathord{\left/
 {\vphantom {M K}} \right.
 \kern-\nulldelimiterspace} K}} \right)$. Then the distance set between each SU and each QU is defined as
 \begin{equation}
 \left\{ {\left. {{d_{k \to m}}} \right|\forall k = 1,2,...,K;\forall m = 1,2,...,M} \right\},
 \end{equation}
 where ${d_{k \to m}}$ representing the relative distance between the $k$-th SU and the $m$-th QU is mathematically defined as
 \begin{equation}
 {d_{k \to m}} \buildrel \Delta \over = {\left\| {{{\bf{w}}_k} - {{\bf{v}}_m}} \right\|_2}.
 \end{equation}
  According to the descending order of the distance set, from the beginning of the maximum distance, if the number of the user pairs that have been constructed with respect to a certain SU is under the defined $L$, then the SU and QU related to the considered distance are paired. On the other hand, if the number of the user pairs that have been constructed with a certain SU is equal to $L$, then this considered distance will be left out and the procedure will be continued. To make it clear, for a certain distance ${{d_{k \to m}}}$ in the set with the descending order, the above explanation can be expressed as
  \begin{equation}
  \left\{\!\! \begin{array}{l}
  \text{if}\;nu{m_k} \!<\! L:\text{Construct\;a\;user\;pair\;with\;}\text{SU}_{k}\text{\;and\;}\text{QU}_{m}\\
  \text{if}\;nu{m_k}\! = \! L:\text{Continue\;with\;the\;descending\;order}
  \end{array} \!\!\right.
  \end{equation}
where $nu{m_k}$ is defined as the number of the user pairs that have been constructed with respect to the $k$-th SU. As a result, multiple distant QUs can be potentially paired with the same SU to construct multiple user pairs, so as to ensure the connectivity of each QU during the flight period when there exist more QUs than SUs. After determining the above specific user pairing scheme, along the trajectory we can schedule the nearest SU in each time slot, which can be mathematically expressed as
  \begin{equation}
  \tilde k = \arg \;\mathop {min}\limits_{k = 1,2,...,K} {d_{UAV \to k}}\left[ n \right],
  \end{equation}
  where ${d_{UAV \to k}}\left[ n \right]$ representing the relative distance between UAV-BS and the $k$-th SU in the $n$-th transmission time slot is mathematically defined as
  \begin{equation}
  {d_{UAV \to k}}\left[ n \right] \buildrel \Delta \over = {\left\| {{\bf{q}}\left[ n \right] - {{\bf{w}}_k}} \right\|_2}.
  \end{equation}
  Then, the paired QU is scheduled according to the user pairing scheme. Moreover, if there exist multiple user pairs with the scheduled SU, the specific scheduled QU in each slot is randomly selected on the basis that each QU has the same number of scheduled time slots.

\subsection{UAV-BS trajectory Design}
According to the specific user-pairing policy among SUs and QUs, the UAV-BS trajectory design is investigated in this section. It should be pointed out that for each user pair, there exists the corresponding optimal horizontal location to provide the best secrecy performance. In such a context, the inherent characteristic of our UAV-BS trajectory design is to sequentially move to these optimal locations with the maximum speed and then hover at the optimal locations for better secrecy performance, which is the typical fly-hover-fly protocol. It is noted that this protocol has been shown as the optimal choice in previous researches. On this basis, there are several remaining problems about the specific UAV-BS trajectory design as follows, consisting of the optimal location design for each user pair, the number of time slots for hovering at each optimal location, and the ordering to reach each optimal location.

\begin{figure}[tbp]
\setlength{\abovecaptionskip}{0.cm}
\setlength{\belowcaptionskip}{-0.cm}
\centering
\includegraphics[width=9cm,height=5cm]{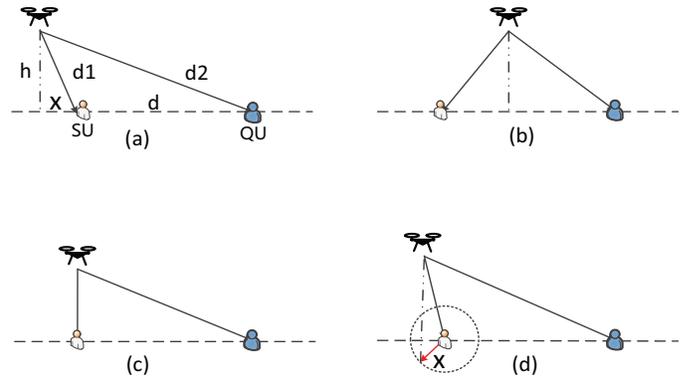}
\caption{The possible optimal hovering locations of UAV-BS for the typical user pair.}\label{hovering_position_model}
\end{figure}

For each user pair, the possible optimal hovering locations of UAV-BS are highly related to the locations of the scheduled users as shown in Fig. 2, where $d$ denotes the distance between the scheduled users, ${d_1}$ and ${d_2}$ represent the distances of the scheduled SU and QU away from UAV-BS, respectively, and $x$ is the distance between the horizontal projection of UAV-BS and the scheduled SU. Due to the priority of security requirements, it is promising to enhance the channel superiority of the scheduled SU compared with the scheduled QU. Therefore, it is clear that the hovering location in case ($b$) is worse than that in case ($c$) due to the smaller $d_2$ and larger $d_1$. Moreover, since ${d_1}$ in case ($d$) equals to that in case ($a$) while ${d_2}$ in case ($d$) is smaller than that in case ($a$), the hovering location in case ($d$) is also worse than that in case ($a$). Then for cases ($a$) and ($c$), we define the following secrecy-rate related function
\begin{equation}\label{e39}
{\rm{f}}\left( x \right) \buildrel \Delta \over = \frac{1}{{{h^2} + {x^2}}} - \frac{1}{{{h^2} + {{\left( {x + d} \right)}^2}}},
\end{equation}
and the first-order derivative with respect to $x$ for $x \ge 0$ is derived at the top of the next page.
\begin{figure*}[ht]
\begin{equation}\label{e40}
\frac{{d{\rm{f}}\left( x \right)}}{{dx}} = \frac{{ - 6d{x^4} - 12{d^2}{x^3} - \left( {4{h^2}d + 8{d^3}} \right){x^2} - \left( {4{h^2}{d^2} + 2{d^4}} \right)x + 2{h^4}d}}{{{{\left( {{x^4} + 2d{x^3} + \left( {2{h^2} + {d^2}} \right){x^2} + 2{h^2}dx + {h^4} + {h^2}{d^2}} \right)}^2}}}.
\end{equation}
\end{figure*}
It is observed from (\ref{e40}) that the first-order derivative at $x=0$ is positive, and thus the hovering location in case ($c$) is worse than that in case ($a$). Then, the optimal ${x^*}$ can be obtained at ${{d{\rm{f}}\left( x \right)} \mathord{\left/
 {\vphantom {{d{\rm{f}}\left( x \right)} {dx}}} \right.
 \kern-\nulldelimiterspace} {dx}} = 0$ by a bisection search method.

After determining the optimal hovering locations, we aim to design the number of time slots for hovering at each optimal location. According to the limited flight period of UAV-BS, there are two possible conditions for hovering. On one hand, due to the limited maximum speed and the limited flight period of UAV-BS, there is no remaining hovering time at each optimal location. On the other hand, the flight period is enough to satisfy the mobility requirement, and thus the remaining hovering time needs to be effectively allocated for each hovering location. To this regard, due to the inherent requirement of fairness in the objective function of problem (\ref{P1}), the remaining time slots are evenly allocated to optimal locations for hovering.

\begin{remark}
For the above designs of UAV-BS trajectory and user-scheduling, which are also regarded as the initial feasible solution for our proposed iterative algorithm in Section \uppercase\expandafter{\romannumeral3}, it is noted that the UAV-BS trajectory is designed mainly based on the locations of SUs, while the scheduled QU in each transmission time slot is selected as the most distant user from the scheduled SU. As a result, the scheduled SU will have better channel quality than the scheduled QU in each transmission slot, and thus the perfect SIC process can also be reasonably satisfied.
\end{remark}

\begin{remark}
\textnormal{If the determined optimal hovering locations for different user pairs with the same SU are adjacent to each other, it is a better choice in practice to select a typical hovering location for all of these pairs to save the time for movement and the energy consumption. Then, the remaining time slots can be proportionally allocated to hovering locations, according to the number of related user pairs at each hovering location.}
\end{remark}

As for the ordering design, the goal is to reduce the flying time between optimal locations during the flight. With the constant maximum speed, this problem is equivalent to the classic travel salesman problem (TSP).
As a result, the efficient algorithm in \cite{TSP} is exploited to solve this problem, which is omitted in this paper due to the limit of pages.

\subsection{Power Allocation Design}
Based on the obtained user-scheduling and UAV-BS trajectory, we investigate the transmit power allocation design in this section. It is noted that after removing the QoS requirements in the original optimization problem and exploit the relationship ${\alpha _1} + {\alpha _2} = 1$, the power allocation problem in each time slot can be expressed as
\begin{subequations}\label{P6}
\begin{align}
&\mathop {\max }\limits_{{\mu},{\alpha _1} } \;\;\;\;\;{\log _2}\left( {1 + \rho {h_k}{\alpha _1}} \right) - \mu    \label{P6e1}\\
&\;\;{\rm{s}}.{\rm{t}}.\;\;\;\;\;\;\mu  \ge {f_1}\left( {{\alpha _1}} \right) \buildrel \Delta \over = {\log _2}\left( {1 + \rho {g_{{m^*}}}{\alpha _1}} \right)  \label{P6e2} \\
&\;\;\;\;\;\;\;\;\;\;\;\;\mu  \ge {\log _2}\left( {1 + \frac{{\rho {g_m}{\alpha _1}}}{{\rho {g_m}\left( {1 - {\alpha _1}} \right) + 1}}} \right), \forall m \ne {m^*}  \label{P6e3} \\
&\;\;\;\;\;\;\;\;\;\;\;\;0 \le {\alpha _1} \le 1, \label{P6e4}
\end{align}
\end{subequations}
which can be seen as a single-variable optimization problem with respect to ${\alpha _1}$, while ${\mu}$ is the introduced slack variable for brevity. Due to the fact that among the unscheduled QUs, the one with the best channel quality will cause the most serious eavesdropping threat to the confidential transmission, and thus constraint (\ref{P6e3}) is equivalently transformed as
\begin{equation}\label{e36}
\mu  \ge {f_2}\left( {{\alpha _1}} \right) \buildrel \Delta \over = {\log _2}\left( {1 + \frac{{\rho {g_{\tilde m}}{\alpha _1}}}{{\rho {g_{\tilde m}}\left( {1 - {\alpha _1}} \right) + 1}}} \right),
\end{equation}
where $\tilde m \buildrel \Delta \over = arg\;ma{x_{m \ne {m^*}}}{g_m}$. On this basis, it is observed that the optimal $\mu$ is highly related to ${\alpha _1}$, and thus by comparing constraints (\ref{P6e2}) and (\ref{e36}), we have
\begin{equation}\label{e37}
\left\{\!\! \begin{array}{l}
{\bf{C1}}:{f_1}\left( {{\alpha _1}} \right) \!\ge\! {f_2}\left( {{\alpha _1}} \right),\;{\alpha _1} \le 1 \!-\! {{\left( {{g_{\tilde m}} \!-\! {g_{{m^*}}}} \right)} \mathord{\left/
 {\vphantom {{\left( {{g_{\tilde m}} - {g_{{m^*}}}} \right)} {\rho {g_{\tilde m}}{g_{{m^*}}}}}} \right.
 \kern-\nulldelimiterspace} {\rho {g_{\tilde m}}{g_{{m^*}}}}}\\
{\bf{C}}2:{f_1}\left( {{\alpha _1}} \right) \!\le\! {f_2}\left( {{\alpha _1}} \right),\;{\alpha _1} \ge 1\! -\! {{\left( {{g_{\tilde m}} \!-\! {g_{{m^*}}}} \right)} \mathord{\left/
 {\vphantom {{\left( {{g_{\tilde m}} - {g_{{m^*}}}} \right)} {\rho {g_{\tilde m}}{g_{{m^*}}}}}} \right.
 \kern-\nulldelimiterspace} {\rho {g_{\tilde m}}{g_{{m^*}}}}}
\end{array} \!\!\right.
\end{equation}

Under the condition ${\bf{C1}}$, which corresponds to the condition that the scheduled QU causes the greatest eavesdropping threat among QUs, the power allocation problem is reformulated as
\begin{subequations}\label{P7}
\begin{align}
&\mathop {\max }\limits_{{\alpha _1} } \;\;\;\;\;{\log _2}\left( {1 + \rho {h_k}{\alpha _1}} \right) - {\log _2}\left( {1 + \rho {g_{{m^*}}}{\alpha _1}} \right)    \label{P7e1}\\
&\;\;{\rm{s}}.{\rm{t}}.\;\;\;\;\;\;0 \le {\alpha _1} \le 1 - {\left\{ {{{\left( {{g_{\tilde m}} - {g_{{m^*}}}} \right)} \mathord{\left/
 {\vphantom {{\left( {{g_{\tilde m}} - {g_{{m^*}}}} \right)} {\rho {g_{\tilde m}}{g_{{m^*}}}}}} \right.
 \kern-\nulldelimiterspace} {\rho {g_{\tilde m}}{g_{{m^*}}}}}} \right\}^ + }, \label{P7e2}
\end{align}
\end{subequations}
where ${\left\{  \cdot  \right\}^ + }$ is introduced to cope with the scenario that the scheduled QU has the best channel quality in the considered time slot. Then, the optimal $\alpha _1^*$ of problem (\ref{P7}) can be derived in the following cases:
\begin{enumerate}[1)]
\item {\bf{Case 1}}: The channel quality of the scheduled QU is better than that of the scheduled SU. In such a context, the objective function of problem (\ref{P7}) is monotonic decreasing with respect to ${\alpha _1}$, and thus the optimal $\alpha _1^* = 0$, which means that the total transmit power is exploited to satisfy the QoS requirement of the scheduled QU in this time slot.

\item {\bf{Case 2}}: The channel quality of the scheduled SU is better than the channel quality of the scheduled QU, and the scheduled QU has the best channel quality among QUs. Under this condition, the objective function is monotonic increasing with respect to ${\alpha _1}$ and constraint (\ref{P7e2}) equals to $0 \le {\alpha _1} \le 1$. As a result, the optimal $\alpha _1^* = 1$, which means that the total transmit power is exploited for confidential transmissions of the scheduled SU.

\item {\bf{Case 3}}: The channel quality of the scheduled SU is better than the channel quality of the scheduled QU, while the unscheduled QU with the greatest eavesdropping threat has better channel quality than the scheduled QU. Then, according to the monotonic increasing objective function, the optimal ${\alpha _1}$ is selected as its upper-bound ${{\left( {{g_{\tilde m}} - {g_{{m^*}}}} \right)} \mathord{\left/
 {\vphantom {{\left( {{g_{\tilde m}} - {g_{{m^*}}}} \right)} {\rho {g_{\tilde m}}{g_{{m^*}}}}}} \right.
 \kern-\nulldelimiterspace} {\rho {g_{\tilde m}}{g_{{m^*}}}}}$. It is observed that the expected signal of the scheduled QU is equivalently recognized as the interference to reduce the eavesdropping ability of the unscheduled QUs, and thus improve the secrecy performance.

\end{enumerate}

Under the condition ${\bf{C2}}$, which corresponds to the condition that the unscheduled QU with the best channel quality causes the greatest eavesdropping threat among QUs, the power allocation problem is reformulated as
\begin{subequations}\label{P8}
\begin{align}
&\mathop {\max }\limits_{{\alpha _1} } \;\;\;{\log _2}\left( {1 + \rho {h_k}{\alpha _1}} \right) + {\log _2}\left( {1 + \rho {g_{\tilde m}} - \rho {g_{\tilde m}}{\alpha _1}} \right)    \label{P8e1}\\
&\;\;{\rm{s}}.{\rm{t}}.\;\;\;\;1 - {{\left( {{g_{\tilde m}} - {g_{{m^*}}}} \right)} \mathord{\left/
 {\vphantom {{\left( {{g_{\tilde m}} - {g_{{m^*}}}} \right)} {\rho {g_{\tilde m}}{g_{{m^*}}}}}} \right.
 \kern-\nulldelimiterspace} {\rho {g_{\tilde m}}{g_{{m^*}}}}} \le {\alpha _1} \le 1, \label{P8e2}
\end{align}
\end{subequations}
where the constant term $ - {\log _2}\left( {1 + \rho {g_{\tilde m}}} \right)$ is left out in the objective function. Moreover, after removing the non-decreasing logarithmic function, it is observed that the objective function is concave with respect to ${\alpha _1}$, and thus through the first-order derivative the optimal point of ${\alpha _1}$ is derived as
\begin{equation}\label{e38}
{{\tilde \alpha }_1} = {{\left( {\rho {h_k}\left( {1 + \rho {g_{\tilde m}}} \right) - \rho {g_{\tilde m}}} \right)} \mathord{\left/
 {\vphantom {{\left( {\rho {h_k}\left( {1 + \rho {g_{\tilde m}}} \right) - \rho {g_{\tilde m}}} \right)} {2{\rho ^2}}}} \right.
 \kern-\nulldelimiterspace} {2{\rho ^2}}}{h_k}{g_{\tilde m}},
\end{equation}
Then, the optimal $\alpha _1^*$ of problem (\ref{P8}) can be derived in the following cases:
\begin{enumerate}[1)]
\item {\bf{Case 4}}: If ${{\tilde \alpha }_1} \le 1 - {{\left( {{g_{\tilde m}} - {g_{{m^*}}}} \right)} \mathord{\left/
 {\vphantom {{\left( {{g_{\tilde m}} - {g_{{m^*}}}} \right)} {\rho {g_{\tilde m}}{g_{{m^*}}}}}} \right.
 \kern-\nulldelimiterspace} {\rho {g_{\tilde m}}{g_{{m^*}}}}}$, the objective function is monotonic decreasing with respect to ${\alpha _1}$ in the range of (\ref{P8e2}), and thus the optimal $\alpha _1^* = 1 - {{\left( {{g_{\tilde m}} - {g_{{m^*}}}} \right)} \mathord{\left/
 {\vphantom {{\left( {{g_{\tilde m}} - {g_{{m^*}}}} \right)} {\rho {g_{\tilde m}}{g_{{m^*}}}}}} \right.
 \kern-\nulldelimiterspace} {\rho {g_{\tilde m}}{g_{{m^*}}}}}$. It is pointed out that the eavesdropping rate of the scheduled QU exactly equals to that of the unscheduled QU with the best channel quality, which means that the goal of power allocation in this case is to balance the eavesdropping threats among all QUs.

\item {\bf{Case 5}}: If $1 - {{\left( {{g_{\tilde m}} - {g_{{m^*}}}} \right)} \mathord{\left/
 {\vphantom {{\left( {{g_{\tilde m}} - {g_{{m^*}}}} \right)} {\rho {g_{\tilde m}}{g_{{m^*}}}}}} \right.
 \kern-\nulldelimiterspace} {\rho {g_{\tilde m}}{g_{{m^*}}}}} \le {{\tilde \alpha }_1} \le 1$, the optimal point ${{\tilde \alpha }_1}$ is feasible for problem (\ref{P8}), and thus the optimal $\alpha _1^* = {{\tilde \alpha }_1}$.

\item {\bf{Case 6}}: If ${{\tilde \alpha }_1} \ge 1$, the objective function is monotonic increasing with respect to ${\alpha _1}$ in the range of (\ref{P8e2}), and thus the optimal $\alpha _1^* = 1$. A reasonable explanation for this case is that the gap of channel gains between the scheduled SU and the most threatening QU is particularly large, and thus the additional interference is not efficient to improve the secrecy performance compared with exploiting the total transmit power for confidential transmissions.

\end{enumerate}

\begin{algorithm}[t]
\setstretch{1}
\caption{Low-Complexity Algorithm}\label{A3}
\begin{algorithmic}[1]
\STATE \textbf{Set}: The maximum scheduling number $L = ceil\left( {{M \mathord{\left/
 {\vphantom {M K}} \right.
 \kern-\nulldelimiterspace} K}} \right)$;
\STATE \textbf{Calculate}: The distance between each SU and QU;
\STATE \ \ \ Determine the user pairing according to the descending order of distances;
\STATE \ \ \ The number of paired QUs with each SU cannot exceed $L$;
\STATE \textbf{Calculate}: The optimal hovering location for each user pair according to (\ref{e40}) and bisection search method, the total time slots for mobility of UAV-BS with the maximum speed, and the remaining time slots for hovering $N_H$;
\STATE \textbf{Repeat}:
\STATE \ \ \ Move to the nearest optimal hovering location with the maximum speed;
\STATE \ \ \ Hover with $floor\left( {{{{N_H}} \mathord{\left/
 {\vphantom {{{N_H}} M}} \right.
 \kern-\nulldelimiterspace} M}} \right)$ time slots;
\STATE \textbf{Until}: Return to the final location;
\STATE \textbf{Calculate}: The time-slotted UAV-BS trajectory ${\bf{Q}}$ and the scheduling indicator ${\bf{A}}$ according to the user-scheduling design in Section \uppercase\expandafter{\romannumeral4}-A;
\STATE \textbf{Calculate}: The optimal power allocation ${{\bf{\alpha }}^*}$ of problem (\ref{P6}) in each time slot according to Section \uppercase\expandafter{\romannumeral4}-C.
\STATE \textbf{If}: The QoS requirements can be satisfied, ${\bf{\alpha }} = {{\bf{\alpha }}^*}$;
\STATE \textbf{Else}: Obtain the power allocation on the basis of ${{\bf{\alpha }}^*}$ through bisection search method;
\STATE \textbf{Output}: UAV-BS trajectory ${\bf{Q}}$, user-scheduling indicator ${\bf{A}}$, and power allocation ${{\bf{\alpha }}}$;
\end{algorithmic}
\end{algorithm}

The power allocation policy for each time slot is proposed in the above cases by leaving out the QoS requirements. Given this power allocation policy, the QoS requirements can be simultaneously satisfied on one hand, while the QoS requirements may not be satisfied due to the limited power allocation for the expected signals of QUs, which is actually a more general case. For the latter, the separately designed power allocation policy needs to be adjusted. Since the received SINR of each QU is monotonically increasing with ${\alpha _2}$ according to (\ref{e13}), the bisection search method can be exploited to meet the QoS requirements, and simultaneously improve the secrecy performance as much as possible.

\subsection{Overall Algorithm}
Based on the aforementioned design, the overall low-complexity algorithm is summarized as \emph{Algorithm \ref{A3}}. It is pointed out that through \emph{Algorithm \ref{A3}}, we are able to obtain a feasible solution of problem (\ref{P1}), which is reasonable according to our analyses and thus can be exploited as an initial solution of \emph{Algorithm \ref{A2}} for further iterative optimization.

\begin{remark}
\textnormal{(Computational Complexity Analysis) As for \emph{Algorithm \ref{A3}}, the main computational complexities are resulted from bisection methods to obtain the optimal hovering locations and adjust the power allocation to satisfy the QoS requirements. Therefore, the overall computational complexity of \emph{Algorithm \ref{A3}} is represented as $O\left( M{\eta_4} + N{\eta_5}\right)$}, where ${\eta_4}$ and ${\eta_5}$ denote the numbers of iterations for the dual bisection search processes, respectively.
\end{remark}

\section{Simulation Results}

\begin{figure}[tbp]
\centering
\subfigure[The convergence of \emph{Algorithm \ref{A2}}]{
\includegraphics[width=9cm,height=6cm]{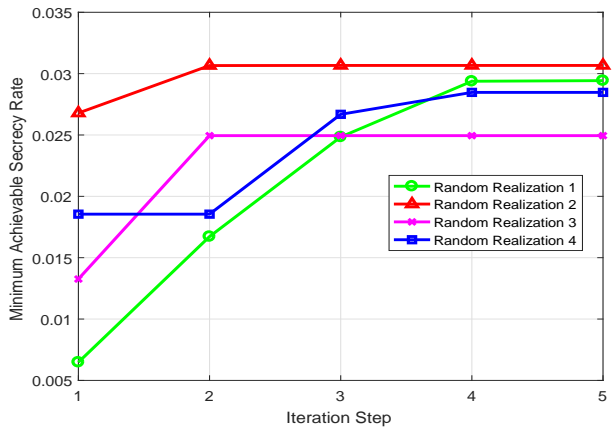}
}
\subfigure[Iteration numbers for different processes]{
\includegraphics[width=9cm,height=6cm]{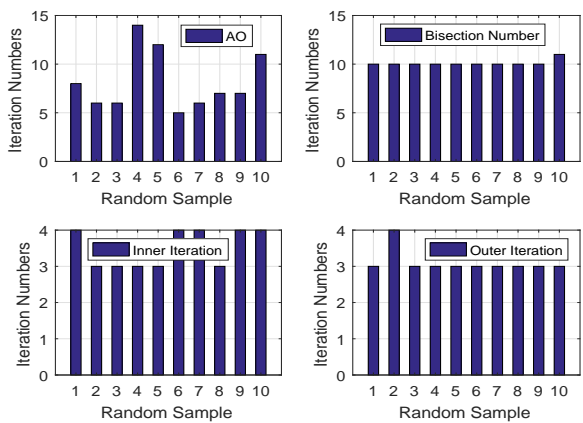}
}
\caption{The convergence performance analysis for different iterative processes with several randomly generated realizations.}\label{IterationSimulation}
\end{figure}

In this section, we numerically evaluate the obtained secrecy performances by exploiting our proposed iterative algorithm and low-complexity algorithm, respectively. In addition, a conventional benchmark OMA scheme with similar optimization process is introduced for performance comparison. Specifically, the limited flight period of UAV-BS is also divided into multiple transmission time slots, while only a SU or QU can be scheduled in each slot. Then, based on the similar optimization process in Section III and IV, since the available transmit power can be exploited for the scheduled user in each time slot, the user-scheduling and UAV-BS trajectory for the proposed benchmark OMA scheme will be jointly designed. It is noted that to satisfy the service requirement of each QU and to improve the achievable secrecy performance of each SU, the initial design for the trajectory of UAV-BS is mainly determined by the locations of both SUs and QUs, instead of only the locations of SUs in the time-slotted NOMA scheme, and the number of time slots for hovering to serve each QU will be determined by their QoS requirements. In addition, the user-scheduling scheme based on the minimum distance principle is adopted. These transmission schemes are recognized as \emph{``Iterative-Optimized NOMA''}, \emph{``Low-Complexity-Designed NOMA''}, and \emph{``Iterative-Optimized OMA''}, respectively, in the following. The simulation parameters are set as following without special instructions. Firstly, the three-dimensional Cartesian coordinate system is constructed, in order to mathematically describe our investigated scenario for simulations. Then, the covered region of UAV-BS can be regarded as a square area in the horizontal plane with a side length of 100m, the center of which is the original point of the constructed coordinate system. Without loss of generality, the initial and final locations of UAV-BS are both set as the original point of the horizontal plane, which aims to provide better service of fairness for the covered area and to periodically charge the UAV-BS. It is assumed that there are $K = 3$ SUs and $M = 3$ QUs and the horizontal locations of which are uniformly distributed in the covered region. As for the parameters setting of UAV-BS, the constant flight height of UAV-BS is set as $H = 100m$, the flight duration of UAV-BS is set as $T = 100s$, the total transmit power of UAV-BS is set as ${{P_{tot}}} = 20 dBm$, and the maximum speed of UAV-BS is set as ${V_{\max }} = 20 m/s$. As for time discretization, the length of each time slot is set as ${\delta _t} = 1s$. As for the parameters setting of legitimate users, the variance of the received noise is ${\sigma^2} =  - 100 dBm$ and the QoS requirement of each QU is set as ${\gamma _m} = 10 bits/Hz$ during the flight period. As for other constant parameters, the channel power gain at the reference distance is ${{\beta _0}} = -70 dB$, the maximum tolerable error and the pre-determined maximum number of steps of both inner and outer iteration processes are set as ${10^{ - 3}}$ and 20, respectively, and the optimization problem is solved for 50 times with randomly distributed user locations.

\begin{figure}[tbp]
\centering
\includegraphics[width=9cm,height=6cm]{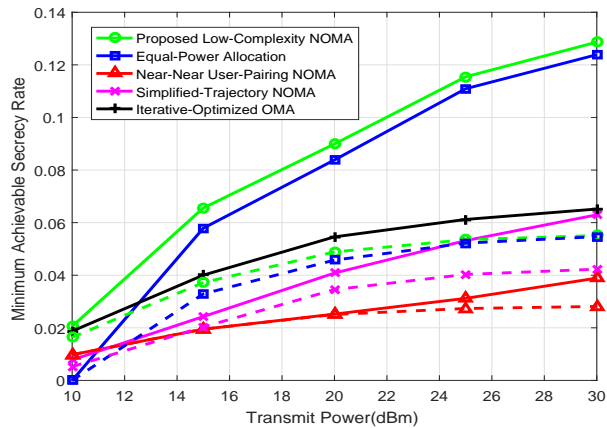}
\caption{Average minimum achievable secrecy rates among SUs versus the available transmit power at UAV-BS, where dash lines represent the obtained performances of low-complexity algorithms and solid lines represent the obtained performances after iterations.}\label{PSimulation}
\end{figure}

To evaluate the convergence performance of our proposed algorithms, Fig. \ref{IterationSimulation}(a) depicts the obtained secrecy performance of each step before convergence by exploiting \emph{Algorithm \ref{A2}} to solve the optimization problem, while Fig. \ref{IterationSimulation}(b) depicts the required numbers of iterations to converge for different iterative processes, all with multiple random realizations. It is observed that all of the realizations can converge to specific values with finite numbers of iterations at least. In addition, it is noticed that the required numbers of iterations to converge with randomly generated user locations does not appear to be much different from each other, which implies that the rate for convergence of our proposed algorithm is stable.

In Fig. \ref{PSimulation}, we investigate the achievable secrecy rate performance versus the total transmit power at UAV-BS. To manifest the performance superiority of our proposed low-complexity algorithm, some benchmarks with comparable complexity are additionally provided: 1) The expected signals of the scheduled SU and QU are allocated with equal power in each transmission time slot. 2) The nearest SU and the nearest QU with respect to UAV-BS in each slot are scheduled. 3) The non-specific UAV-BS trajectory with user locations is adopted, where UAV-BS periodically moves along the sides of the square area above the horizontal plane, determined by points set $\left\{ {\left( {50,50} \right),\left( {50, - 50} \right),\left( { - 50, - 50} \right),\left( { - 50,50} \right)} \right\}$. The above benchmarks are named as \emph{``Equal-Power Allocation''}, \emph{``Near-Near User-Pairing''}, and \emph{``Simplified UAV-BS Trajectory''}, respectively.
Then, the obtained secrecy performances after iterations with different low-complexity benchmarks as initial feasible solutions are also compared.
It is observed that our proposed NOMA transmission scheme after iterative optimization process can achieve better secrecy performance compared with the proposed benchmark schemes, which validates the performance superiority of our proposed transmission scheme.
Then, it is observed that there is only a little performance difference between our proposed NOMA transmission scheme and \emph{``Equal-Power Allocation''} for both low-complexity design and iterative optimization. This observation indicates that if the QoS requirements can be satisfied, the impact of power allocation for each transmission time slot on the resulted secrecy performance is quite limited, especially for high transmit SNR scenario.
Moreover, it is observed that compared with the OMA transmission scheme, the proposed benchmarks with NOMA transmission can even lead to the worse secrecy performance, which demonstrates the significance of the user-scheduling and UAV-BS trajectory design for NOMA transmission.
Finally, it is noted that the performance improvement will gradually decrease with the increase of transmit power, since the achievable secrecy rate will be determined by the ratio of the main channel to the most threatening eavesdropping channel for extremely high transmit SNR, which is independent of the transmit power.

\begin{figure}[tbp]
\centering
\subfigure[SU-Scheduling]{
\includegraphics[width=9cm,height=6cm]{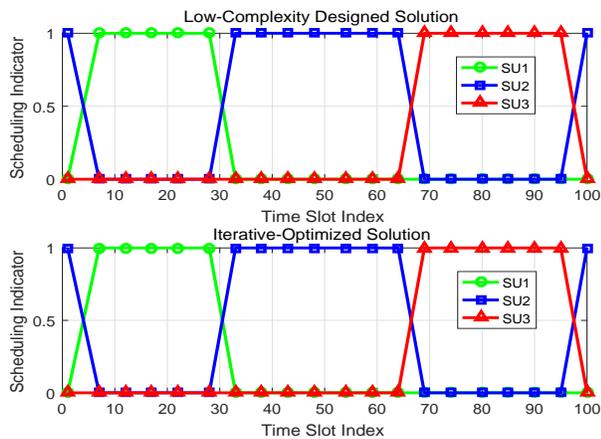}
}
\subfigure[QU-Scheduling]{
\includegraphics[width=9cm,height=6cm]{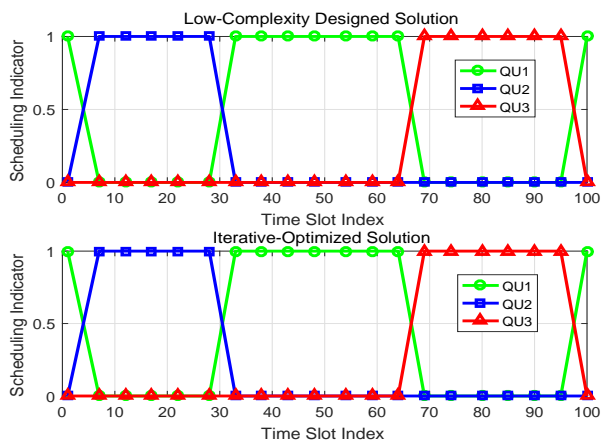}
}
\caption{The Scheduling indicators of SUs and QUs in each time slot for both low-complexity-designed NOMA and iterative-optimized NOMA, respectively.}\label{SchedulingSimulation}
\end{figure}

Fig. \ref{SchedulingSimulation} compares the scheduling of users in each time slot for \emph{low-Complexity-Designed NOMA} and \emph{iterative-Optimized NOMA} for one random realization of user locations. It is observed that the user-scheduling is nearly unchanged after iterative optimization, which demonstrates the rationality of our proposed user-scheduling policy for low-complexity design. Then, together with Fig. \ref{PSimulation}, we can further draw a conclusion that the trajectory design of the iterative optimization has a significant impact on the secrecy performance improvement compared with the obtained secrecy performance by low-complexity algorithm.

\begin{figure}[tbp]
\centering
\subfigure[QoS Requirements]{
\includegraphics[width=9cm,height=6cm]{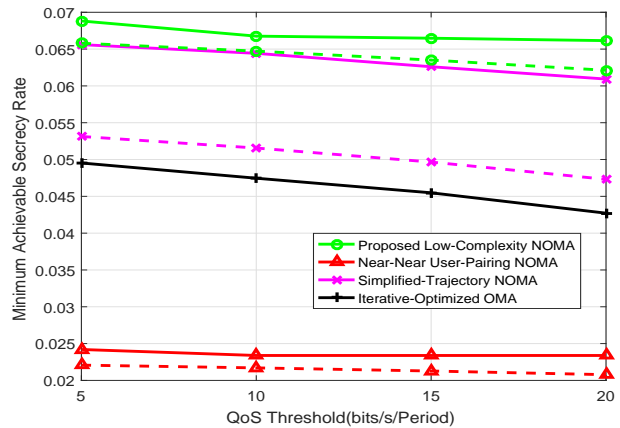}
}
\subfigure[Flight Period]{
\includegraphics[width=9cm,height=6cm]{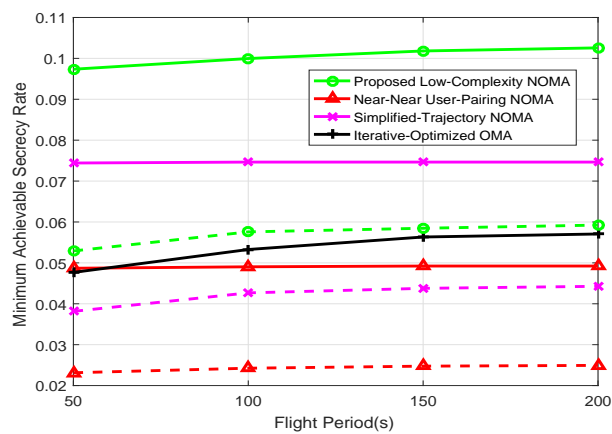}
}
\caption{Average minimum achievable secrecy rates among SUs versus the QoS requirement of each QU and the flight period of UAV-BS, where dash lines represent the obtained performances of low-complexity algorithms and solid lines represent the obtained performances after iterations.}\label{QoSandFlightSimulation}
\end{figure}

\begin{figure*}[htbp]
\setlength{\abovecaptionskip}{0.3cm}
\setlength{\belowcaptionskip}{-0.3cm}
\centering

\subfigure[Reference Scenario]{
\begin{minipage}[t]{0.49\linewidth}
\centering
\includegraphics[width=9cm,height=6cm]{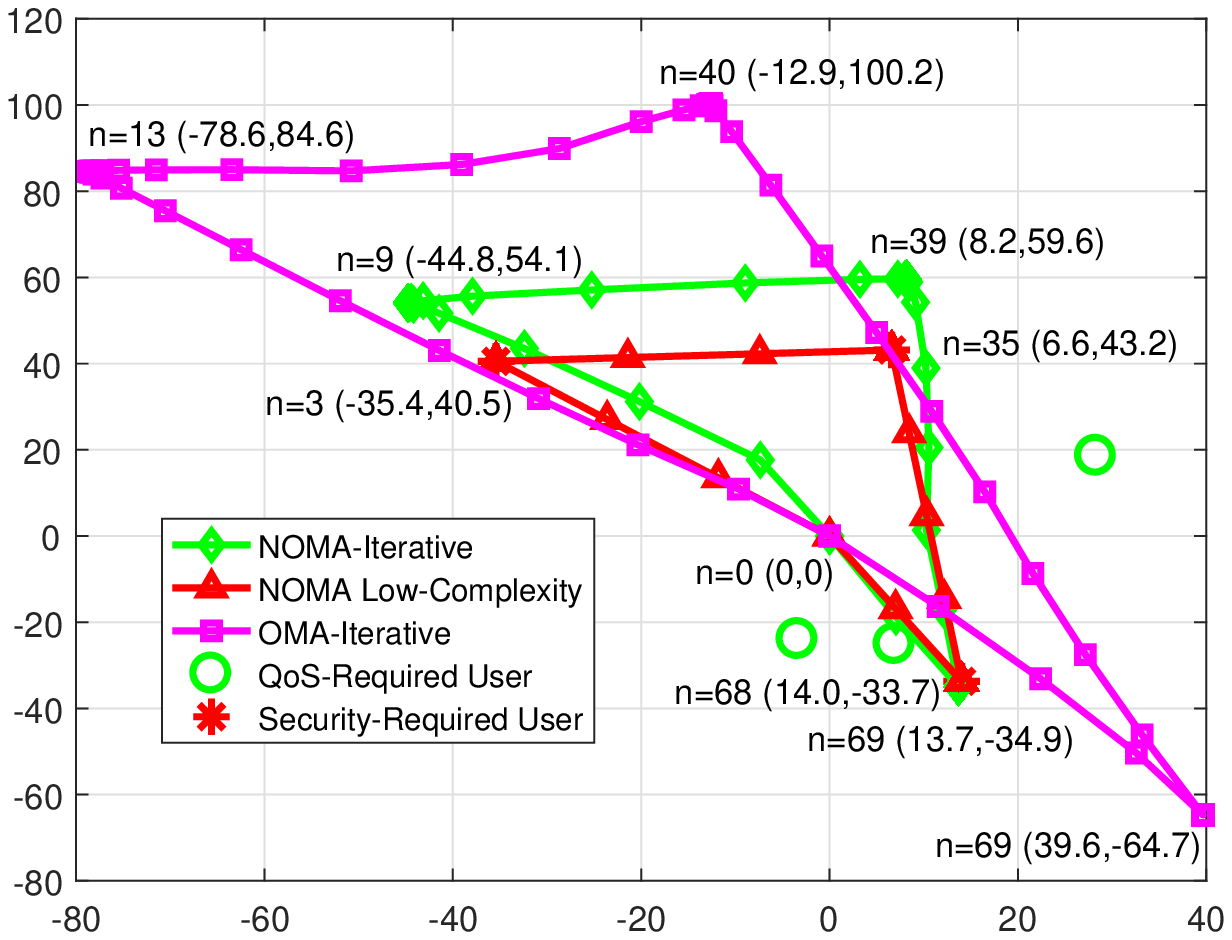}
\end{minipage}%
}%
\subfigure[${{P_{tot}}} = 30 dBm$]{
\begin{minipage}[t]{0.49\linewidth}
\centering
\includegraphics[width=9cm,height=6cm]{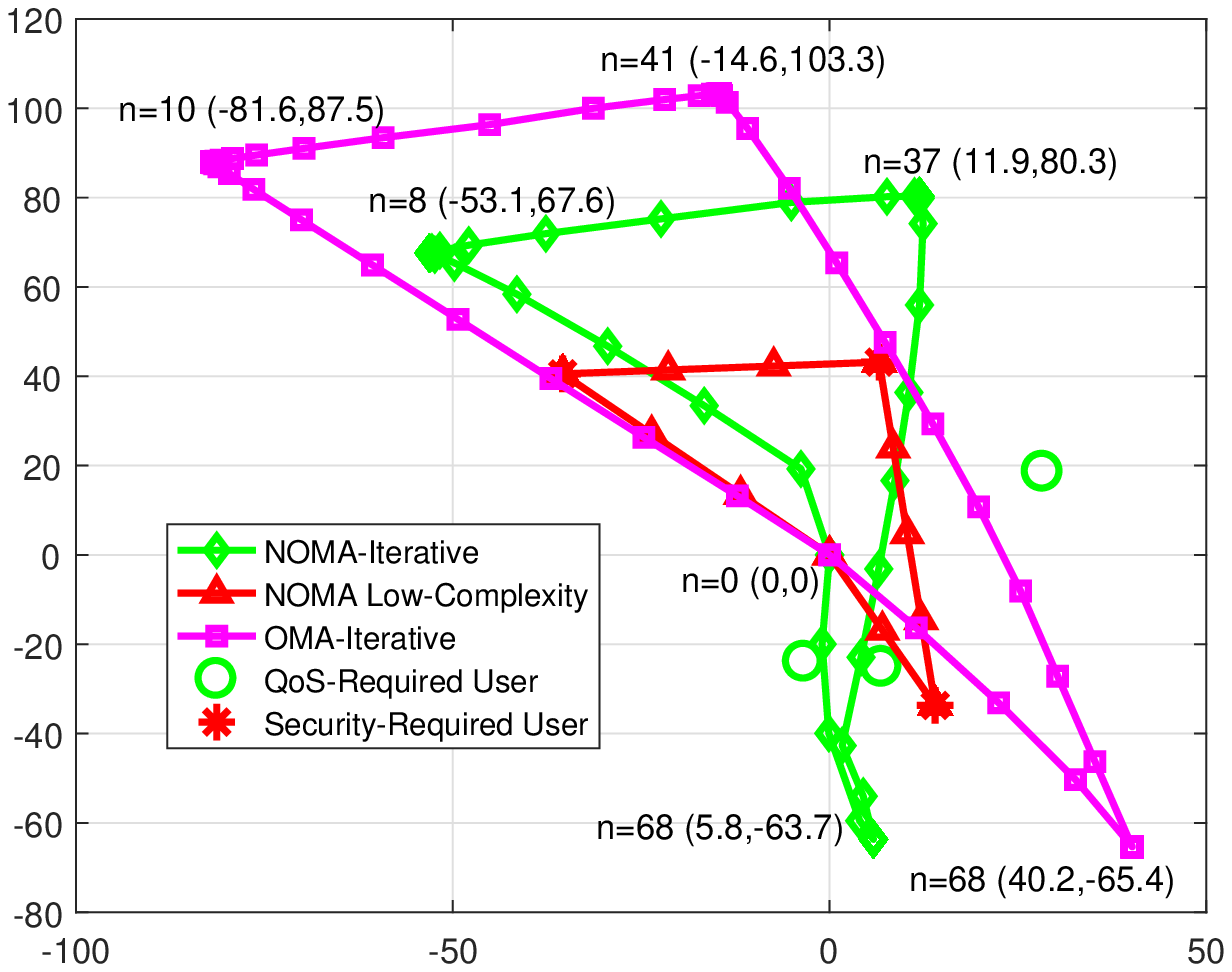}
\end{minipage}%
}%

\subfigure[${\gamma _m} = 20 bits/Hz$]{
\begin{minipage}[t]{0.49\linewidth}
\centering
\includegraphics[width=9cm,height=6cm]{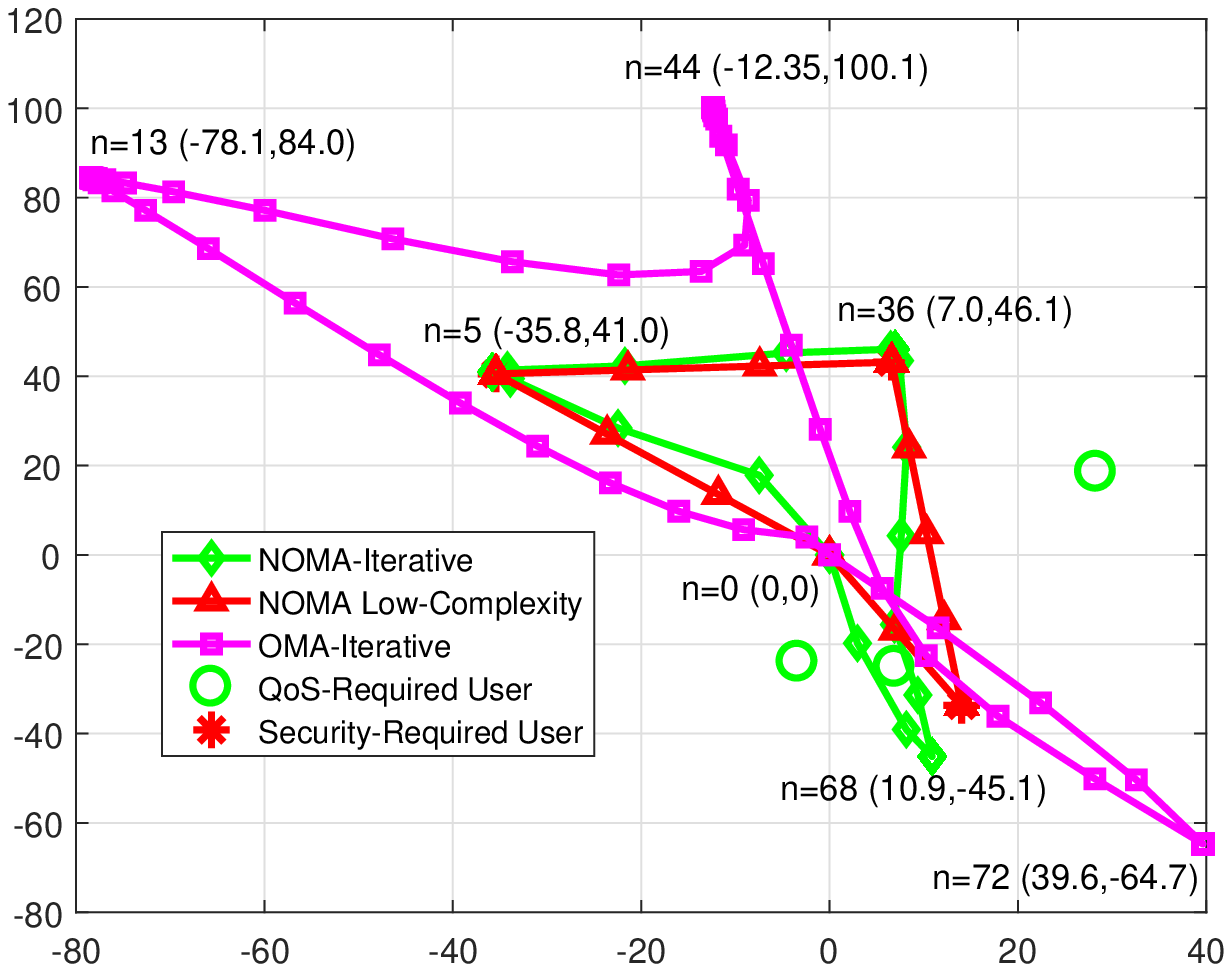}
\end{minipage}
}%
\subfigure[$T = 60s$]{
\begin{minipage}[t]{0.49\linewidth}
\centering
\includegraphics[width=9cm,height=6cm]{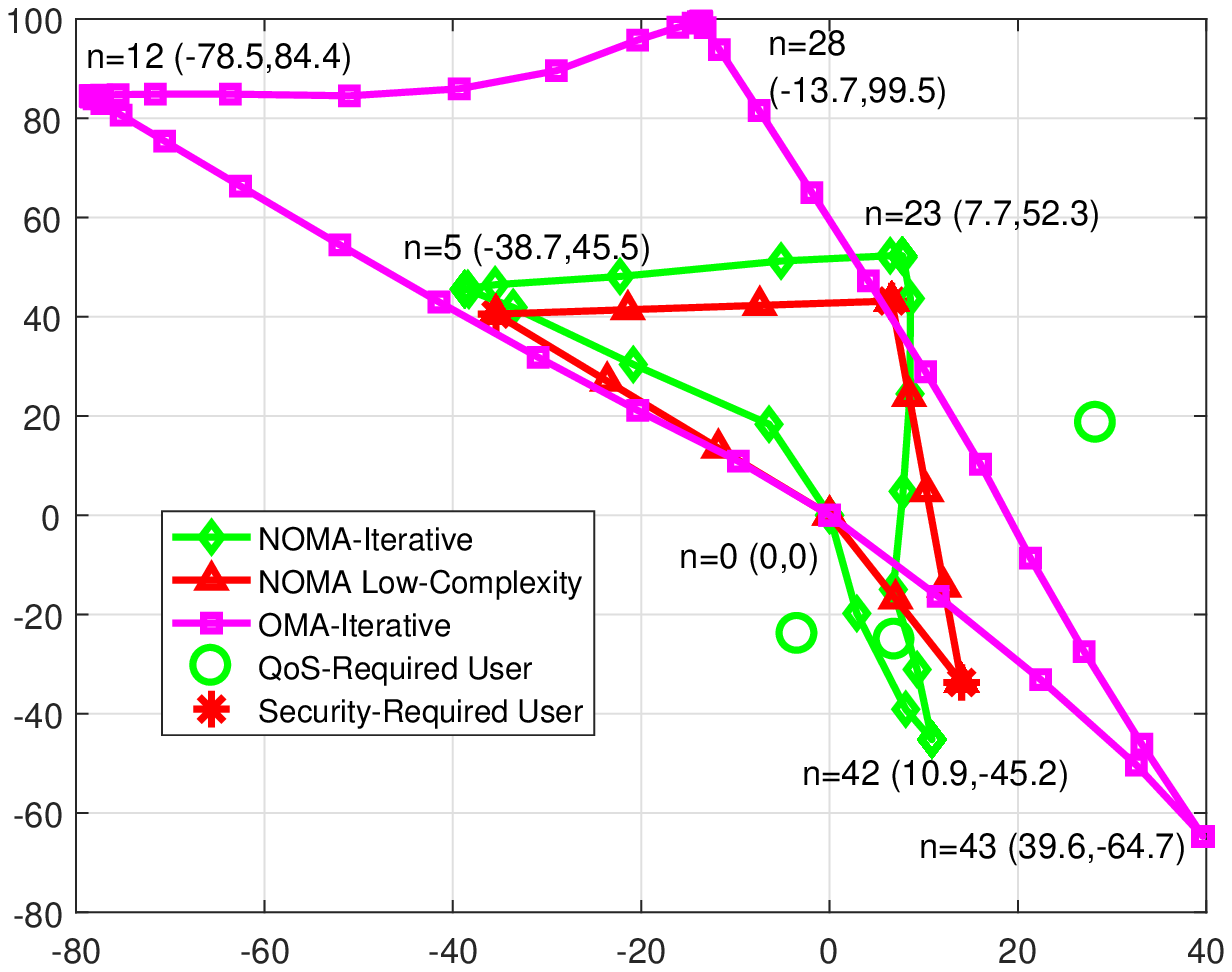}
\end{minipage}
}%

\centering
\caption{Specific UAV-BS trajectories with different parameter settings for each transmission scheme.}\label{TrajSimulation}
\end{figure*}

Then, to evaluate the impact of QoS requirements of QUs and the flight period of UAV-BS on the obtained secrecy performance, Fig. \ref{QoSandFlightSimulation}(a) and Fig. \ref{QoSandFlightSimulation}(b) are provided. It is noted that due to the limit impact of power allocation on the obtained secrecy performance, the \emph{``Equal-Power Allocation''} benchmark is omitted.
From Fig. \ref{QoSandFlightSimulation}(a), it is observed that our proposed NOMA transmission scheme can perform better than other benchmark schemes even with the QoS requirements becoming more stringent.
Moreover, it is also observed that the obtained secrecy performance of NOMA transmission scheme with inappropriate user-scheduling design, such as the \emph{``Near-Near User-Pairing''} scheme, will lead to worse performance compared with the OMA transmission scheme irrespective of the QoS requirements.
From Fig. \ref{QoSandFlightSimulation}(a), though only the limited performance degradation is observed with the increase of QoS requirements, it is reasonable to predict that the achievable secrecy performance will be sharply degraded with higher QoS requirements. This observation can be accounted for that to satisfy the higher QoS requirements, the determined hovering locations of UAV-BS may exist in the neighborhood of QUs, and thus the channel quality of the scheduled SU is degraded, even worse than the scheduled QU, which will significantly deteriorate the secrecy performance.

From Fig. \ref{QoSandFlightSimulation}(b), the similar conclusions with Fig. \ref{QoSandFlightSimulation}(a) can be drawn with respect to the achievable secrecy performances between different schemes, which once more demonstrates the significance of the user-scheduling and UAV-BS trajectory design for NOMA transmission.
In addition, for the \emph{``Near-Near User-Pairing''} and \emph{``Simplified UAV-BS Trajectory''} benchmark schemes, it is noted that there is nearly no improvement of the obtained secrecy performance after iterations with the increase of flight period, and thus the performance gap between these schemes and our proposed NOMA transmission scheme will be enlarged.
Moreover, as the pace of secrecy performance improvement gradually becomes slow with the increase of flight period, we can predict that the obtained achievable secrecy performances of different transmission schemes will finally converge to some stationary values. This is due to the fact that the achievable secrecy performance will be mainly determined by UAV-BS hovering at specific optimal locations at this time, which can be equivalently recognized as multiple static UAV-BS transmission cases and thus the average secrecy performance during the flight period will be unchanged.

Fig. \ref{TrajSimulation} depicts the specific UAV-BS trajectories for one random sample of user locations with different parameter settings for each transmission scheme to reflect the the impact of system parameters on the UAV-BS trajectory design of different transmission schemes.
Firstly, it is observed from Fig. \ref{TrajSimulation}(a) that the obtained UAV-BS trajectory of \emph{low-Complexity-Designed NOMA} and \emph{iterative-Optimized NOMA} have similar characteristics, but this similarity still has a great impact on secrecy performance as previously indicated in Fig. \ref{SchedulingSimulation}. However, the designed trajectory of \emph{iterative-Optimized OMA} is obviously different, in which the total flight distance will be significantly increased. This is due to the fact that there is no equivalent interference signal in each time slot for OMA transmission scheme. At this time, the only way to reduce the threat of potential eavesdroppers is to keep UAV-BS relatively far away from them. In general, it is noted that the obtained UAV-BS trajectories of all the transmission schemes
have the characteristics of hovering at some optimal positions and reaching these optimal position in turn with the maximum speed, from the initial position to the final position.

Compared with Fig. \ref{TrajSimulation}(a), it is observed from Fig. \ref{TrajSimulation}(b) that as the available transmit power increases, the flight distance of UAV-BS will be enlarged, and thus the hovering positions are relatively further away from QUs. At this time, their QoS requirements can still be satisfied while their potential eavesdropping capability will be significantly degraded to achieve better secrecy performance. Then, it is observed from Fig. \ref{TrajSimulation}(c) that as the QoS requirements of QUs become more stringent, UAV-BS will move closer to distant QUs to satisfy their QoS requirements. To this regard, the channel quality of the scheduled SU is highly degraded and can even be worse than the channel quality of the scheduled QU, and thus the achievable secrecy performance will be significantly decreased. Finally, by comparing Fig. \ref{TrajSimulation}(a) and Fig. \ref{TrajSimulation}(d), it is shown that the obtained UAV-BS trajectories for different flight periods of UAV-BS are quite similar if the flight period is enough for UAV-BS to move to the certain hovering locations. At this time, the main difference for these two cases is the number of time slots for hovering, which also significantly affects the achievable secrecy performance.

\section{Conclusion}
In this paper, we have investigated the secrecy performance in a downlink UAV-BS-enabled multi-user NOMA transmission system with legitimate users categorized as security-required and QoS-required users, while these QoS-required users can potentially act as internal eavesdroppers overhearing secrecy transmissions. Then, the optimization problem with respect to user-scheduling, power allocation, and trajectory design has been formulated. To achieve the trade-off between objective performance and computational complexity, efficient iterative-based and low-complexity-based algorithms have been proposed to solve the problem. It has been shown that our proposed NOMA transmission schemes can have sufficient superiority over the conventional OMA benchmark, and the impact of some key system parameters, such as the flight period and the available transmit power of UAV-BS, on the objective performance and UAV-BS trajectory design has been provided.


\begin{thebibliography}{100}
\begin{spacing}{1}
\bibitem{R1}
S. Hayat, E. Yanmaz, and R. Muzaffar, ``Survey on unmanned aerial vehicle networks for civil applications: A communications viewpoint,'' \emph{IEEE Commun. Surv. Tuts.}, vol. 18, no. 4, pp. 2624--2661, Fourth Quarter 2016.

\bibitem{R2}
Y. Zeng, R. Zhang, and T. J. Lim, ``Wireless communications with unmanned aerial vehicles: Opportunities and challenges," \emph{IEEE Commun. Mag.}, vol. 54, no. 5, pp. 36--42, May 2016.

\bibitem{R17}
Y. Zeng, Q. Wu, and R. Zhang, ``Accessing from the sky: A tutorial on UAV communications for 5G and beyond,'' \emph{Proc. IEEE}, vol. 107, no. 12, pp. 2327--2375, Dec. 2019.

\bibitem{R18}
A. A. Khuwaja, Y. Chen, N. Zhao, M.-S. Alouini, and P. Dobbins, ``A survey of channel modeling for UAV communications," \emph{IEEE Commun. Surv. Tuts.}, vol. 20, no. 4, pp. 2804--2821, Fourth Quarter 2018.

\bibitem{R20}
Y. Liu, Z. Qin, M. Elkashlan, Z. Ding, A. Nallanathan, and L. Hanzo, ``Nonorthogonal multiple access for 5G and beyond,'' \emph{Proc. IEEE}, vol. 105, no. 12, pp. 2347--2381, Dec. 2017.

\bibitem{R11}
Y. Liu, Z. Qin, Y. Cai, Y. Gao, G. Y. Li, and A. Nallanathan, ``UAV communications based on non-orthogonal multiple access," \emph{IEEE Wireless Commun.}, vol. 26, no. 1, pp. 52--57, Feb. 2019.

\bibitem{R12}
W. Mei and R. Zhang, ``Uplink cooperative NOMA for cellular-connected UAV," \emph{IEEE J. Sel. Topics Signal Process.}, vol. 13, no. 3, pp. 644--656, Jun. 2019.

\bibitem{R13}
T. Hou, Y. Liu, Z. Song, X. Sun, and Y. Chen, ``Multiple antenna aided NOMA in UAV networks: A stochastic geometry approach," \emph{IEEE Wireless Commun.}, vol. 67, no. 2, pp. 1031--1044, Feb. 2019.

\bibitem{R32}
T. Hou, Y. Liu, Z. Song, X. Sun, and Y. Chen, ``Exploiting NOMA for UAV communications in large-scale cellular networks," \emph{IEEE Trans. Commun.}, vol. 67, no. 10, pp. 6897--6911, Jul. 2019.

\bibitem{R14}
J. Sun, Z. Wang, and Q. Huang, ``Cyclical NOMA based UAV-enabled wireless network," \emph{IEEE Access}, vol. 7, pp. 4248--4259, Dec. 2018.

\bibitem{R33}
N. Zhao, X. Pang, Z. Li, Y. Chen, F. Li, Z. Ding, and M.-S. Alouini, ``Joint trajectory and precoding optimization for UAV-assisted NOMA networks," \emph{IEEE Trans. Commun.}, vol. 67, no. 5, pp. 3723--3735, Jan. 2019.

\bibitem{R34}
F. Cui, Y. Cai, Z. Qin, M. Zhao, and G. Y. Li, ``Multiple access for mobile-UAV networks: Joint trajectory design and resource allocation," \emph{IEEE Trans. Commun.}, vol. 67, no. 7, pp. 4980--4994, Jul. 2019.

\bibitem{R3}
H.-M. Wang, X. Zhang, and J.-C. Jiang, ``UAV-involved wireless physical-layer secure communications: Overview and research directions," \emph{IEEE Wireless Commun.}, vol. 26, no. 5, pp. 32--39, Oct. 2019.

\bibitem{RPLS}
M. Bloch and J. Barros, \emph{Physical-layer security: From information theory to secure engineering.} Cambridge University Press, Aug. 2008.

\bibitem{RPLS2}
B. He, X. Zhou, and T. D. Abhayapala, ``Wireless physical layer security
with imperfect channel state information: A survey,'' \emph{ZTE Commun.}, vol.
11, no. 3, pp. 11--19, Sept. 2013.

\bibitem{RPLS1}
K. Huang, L. Jin, and Z. Zhong, ``5G physical layer security technology: Enhancing security by communication,'' \emph{ZTE Technol. J.}, vol. 25, no. 4, pp. 43--49, Aug. 2019.

\bibitem{R4}
G. Zhang, Q. Wu, M. Cui, and R. Zhang, ``Securing UAV communications via joint trajectory and power control,'' \emph{IEEE Trans. Wireless Commun.}, vol. 18, no. 2, pp. 1376--1389, Jan. 2019.

\bibitem{R23}
M. Cui, G. Zhang, Q. Qu, and D. W. K. Ng, ``Robust trajectory and transmit power design for secure UAV communications,'' \emph{IEEE Trans. Veh. Technol.}, vol. 67, no. 9, pp. 9042--9046, Jun. 2018.

\bibitem{R24}
Z. Li, M. Chen, C. Pan, N. Huang, Z. Yang, and A. Nallanathan, ``Joint trajectory and communication design for secure UAV networks,'' \emph{IEEE Commun. Lett.}, vol. 23, no. 4, pp. 636--639, Feb. 2019.


\bibitem{R6}
Q. Wang, Z. Chen, W. Mei, and J. Fang, ``Improving physical layer security using UAV-enabled mobile relaying," \emph{IEEE Wireless Commun. Lett.}, vol. 6, no. 3, pp. 310--313, Jun. 2017.

\bibitem{R7a}
Q. Wang, Z. Chen, H. Li, and S. Li, ``Joint power and trajectory design for physical-layer secrecy in the UAV-aided mobile relaying system," \emph{IEEE Access}, vol. 6, pp. 62849--62855, 2018.

\bibitem{R8}
A. Li, Q. Wu, and R. Zhang, ``UAV-enabled cooperative jamming for improving secrecy of ground wiretap channel," \emph{IEEE Wireless Commun. Lett.}, vol. 8, no. 1, pp. 181--184, Feb. 2019.

\bibitem{R9}
Y. Zhou, P. L. Yeoh, H. Chen, Y. Li, R. Schober, L. Zhuo, and B. Vucetic, ``Improving physical layer security via a UAV friendly jammer for unknown eavesdropper location," \emph{IEEE Trans. Veh. Technol.}, vol. 67, no. 11, pp. 11280--11284, Nov. 2018.

\bibitem{R25}
Y. Zhu, G. Zheng, and M. Fitch, ``Secrecy rate analysis of UAV-enabled mmWave networks using mat\'{e}rn hardcore point processes," \emph{IEEE J. Sel. Areas Commun.}, vol. 36, no. 7, pp. 1397--1409, Jul. 2018.

\bibitem{R26}
J. Yao and J. Xu, ``Secrecy transmission in large-scale UAV-enabled wireless networks," \emph{IEEE Trans. Commun.}, vol. 67, no. 11, pp. 7656--7671, Aug. 2019.

\bibitem{R19}
Y. Cai, F. Cui, Q. Shi, M. Zhao, and G. Li, ``Dual-UAV-enabled secure communications: Joint trajectory design and user scheduling," \emph{IEEE J. Sel. Areas Commun.}, vol. 36, no. 9, pp. 1972--1985, Aug. 2018.

\bibitem{R27}
H. Lee, S. Eom, J. Park, and I. Lee, ``UAV-aided secure communications with cooperative jamming," \emph{IEEE Trans. Veh. Technol.}, vol. 67, no. 10, pp. 9385--9392, Oct. 2018.

\bibitem{R28}
C. Zhong, J. Yao, and J. Xu, ``Secure UAV communication with cooperative jamming and trajectory control," \emph{IEEE Commun. Lett.}, vol. 23, no. 2, pp. 286--289, Dec. 2018.

\bibitem{R29}
X. Zhou, Q. Wu, S. Yan, F. Shu, and J. Li, ``UAV-enabled secure communications: joint trajectory and transmit power optimization," \emph{IEEE Trans. Veh. Technol.}, vol. 68, no. 4, pp. 4069--4073, Feb. 2019.


\bibitem{RNOMAPLS}
Y. Zhang, H.-M. Wang, Q. Yang, and Z. Ding, ``Secrecy sum rate maximization in non-orthogonal multiple access,'' \emph{IEEE Commun. Lett.}, vol. 20, no. 5, pp. 930--933, May 2016.

\bibitem{R15}
N. Rupasinghe, Y. Yapici, I. Guvenc, H. Dai, and A. Bhuyan, ``Enhancing physical layer security for NOMA transmission in mmWave drone networks," in \emph{Proc. IEEE 52nd Asilomar Conf. Signals Syst. Comput.}, Oct. 2018.

\bibitem{R30}
X. Sun, W. Yang, and Y. Cai, ``Secure communication in NOMA assisted millimeter wave SWIPT UAV networks," \emph{IEEE Internet Things J.}, vol. 7, no. 3, pp. 1884--1897, Mar. 2020.


\bibitem{R32a}
N. Zhao, Y. Li, Y. Chen, W. Lu, J. Wang, and X. Wang, ``Security enhancement for NOMA-UAV networks,'' \emph{IEEE Trans. Veh. Technol.}, 2020, to be published.

\bibitem{R33a}
Z. Yin, M. Jia, W. Wang, N. Cheng, F. Lyu, and X. Shen, ``Max-min secrecy rate for NOMA-based UAV-assisted communications with protected zone,'' in \emph{Proc. Global. Commun. Conf. (GLOBECOM)}, Waikoloa, HI, USA, Dec. 2019.

\bibitem{R21}
H.-M. Wang, X. Zhang, Q. Yang, and T. A. Tsiftsis, ``Secure users oriented downlink MISO NOMA,'' \emph{IEEE J. Sel. Topics Signal Process.}, vol. 13, no. 3, pp. 671--684, Jun. 2019.

\bibitem{Rfootnote}
Q. Wu, Y. Zeng, and R. Zhang, ``Joint trajectory and communication design for multi-UAV enabled wireless networks,''. \emph{IEEE Trans. Wireless Commun.}, vol. 17, no. 3, pp. 2109--2121, Mar. 2018.

\bibitem{R35}
Y. Li, M. Jiang, Q. Zhang, Q. Li, and J. Qin, ``Secure beamforming in downlink MISO nonorthogonal multiple access systems," \emph{IEEE Trans. Veh. Technol.}, vol. 66, no. 8, pp. 7563--7567, Aug. 2017.

\bibitem{R36}
Q. Wu and R. Zhang, ``Common throughput maximization in UAV-enabled OFDMA systems with delay consideration,'' \emph{IEEE Trans. Commun.}, vol. 66, no. 12, pp. 6614--6627, Dec. 2018.

\bibitem{R37}
K.-Y. Wang, A. M.-C. So, T.-H. Chang, W.-K. Ma, and C.-Y Chi, ``Outage constrained robust transmit optimization for multiuser MISO downlinks: Tractable approximations by conic optimization,'' \emph{IEEE Trans. Signal Process.}, vol. 62, no. 21, pp. 5690--5705, Nov. 2014.

\bibitem{TSP}
M. Dorigo and L. M. Gambardella, ``Ant colony system: A cooperative learning approach to the traveling salesman problem,'' \emph{IEEE Trans. Evol. Comput.}, vol. 1, no. 1, pp. 53--66, Apr. 1997.

\end{spacing}
\end{thebibliography}
\end{document}